\documentclass[review]{svjour3}

\usepackage[english]{babel}
\usepackage[utf8x]{inputenc}
\usepackage[T1]{fontenc}
\usepackage{amssymb}
\usepackage{amsmath}
\usepackage{graphicx,txfonts}
\usepackage[colorinlistoftodos]{todonotes}

\usepackage{multirow}

\usepackage{hyperref}
\usepackage{url}

\usepackage{caption}
\usepackage{subcaption}
\usepackage{pgfplots}

\emergencystretch=\maxdimen
\usepgfplotslibrary{external} 
\begin{document}

\title{\textsc{TextBenDS}: a generic Textual data Benchmark for Distributed Systems}

\author{Ciprian-Octavian Truică \and 
        Elena Apostol \and
        Jérôme Darmont \and
        Ira Assent
}

\institute{ 
    Ciprian-Octavian Truică \at
        Computer Science and Engineering Department, Faculty of Automatic Control and Computers, University Politehnica of Bucharest, Bucharest, Romania
        \email{ciprian.truica@cs.pub.ro} \\
        Department of Computer Science, Aarhus University, Aarhus, Denmark
        \email{ciprian.truica@cs.au.dk}
    \and
    Elena Apostol \at
        Computer Science and Engineering Department, Faculty of Automatic Control and Computers, University Politehnica of Bucharest, Bucharest, Romania
        \email{elena.apostol@cs.pub.ro}
    \and 
    Jérôme Darmont \at
        Université de Lyon, Lyon 2, ERIC EA 3083, France
        \email{jerome.darmont@univ-lyon2.fr}
    \and 
    Ira Assent \at
        DIGIT, Department of Computer Science, Aarhus University, Denmark 
        \email{ira@cs.au.dk}
}

\date{Received: date / Accepted: date}

\maketitle

\begin{abstract}
    Extracting top-$k$ keywords and documents using weighting schemes are popular techniques employed in text mining and machine learning for different analysis and retrieval tasks. 
    The weights are usually computed in the data preprocessing step, as they are costly to update and keep track of all the modifications performed on the dataset.
    Furthermore, computation errors are introduced when analyzing only subsets of the dataset.
    Therefore, in a Big Data context, it is crucial to lower the runtime of computing weighting schemes, without hindering the analysis process and the accuracy of the machine learning algorithms.
    To address this requirement for the task of top-$k$ keywords and documents, it is customary to design benchmarks that compare weighting schemes within various configurations of distributed frameworks and database management systems. 
    Thus, we propose a generic document-oriented benchmark for storing textual data and constructing weighting schemes (\textsc{TextBenDS}). 
    Our benchmark offers a generic data model designed with a multidimensional approach for storing text documents. 
    We also propose using aggregation queries with various complexities and selectivities for constructing term weighting schemes, that are utilized in extracting top-$k$ keywords and documents. 
    We evaluate the computing performance of the queries on several distributed environments set within the Apache Hadoop ecosystem.
    Our experimental results provide interesting insights. As an example, MongoDB proves to have the best overall performance, while Spark's execution time remains almost the same, regardless of the weighting schemes. 
\end{abstract}


\section{Introduction}\label{sec:introduction}

With the increase in generated textual data and the many challenges related to extracting knowledge and patterns from large amounts of textual documents, new methods that use dynamic text processing need to be employed. Among the techniques applied in several domains such as text and opinion mining, machine learning and information retrieval, top-$k$ keywords and documents extraction are very frequently used~\cite{Paltoglou2010,SparckJones2000a,SparckJones2000b}. 

Weighting keywords and extracting the top-$k$ frequent terms from a corpus is successfully employed in document clustering~\cite{Yin2018,Wang2017} where they are used as metrics to compute the inter-cluster distances. In topic modeling~\cite{Krasnashchok2018,Truica2016}, term weighting schemes are used to measure the relevance of a word to a topic. For event detection~\cite{Guille2015}, term weights are used to measure the appearance of bursty topic in Online Social Networks. Trend discovery~\cite{Bouakkaz2016,Bringay2011,Ravat2008} employs Text Cubing and Online Analytical Processing (OLAP)~\cite{Zhang2009,Zhang2012} to construct term weighting schemes, which are used for analyzing the impact of products in Social Media. Sentiment analysis uses weighting schemes to vectorize textual data before detecting the polarity of each document~\cite{Paltoglou2010}.

Finding the top-$k$ documents that are most similar to a query is one of the core tasks in information retrieval~\cite{Hofmann2017} and natural language processing~\cite{Lavrenko2017}. In information retrieval, ranking functions are used to score the similarity between a search query and a corpus of documents and return to the user the most relevant documents for their search. Ranking functions are also used in Natural Language Processing for the task of finding hidden semantic structures in textual data using techniques such as Latent Semantic Indexing~\cite{Deerwester1990}.

Considering this context, choosing a processing framework is not easy. Thus, benchmarking is usually used to compare combinations of weighting schemes, computing strategies and physical implementations. However, most big data benchmarks focus on MapReduce operations and do not specifically target textual data~\cite{Huang2010,tpcxhs15,Wang2014}. Moreover, the few benchmarks that do feature textual data~\cite{Ferrarons2014,Gattiker2013} are still at the methodology or specification stages. 

T$^2$K$^2$D$^2$ (Twitter Top-$k$ Keywords and Documents multidimensional) benchmark~\cite{Truica2018} solves some of the current issues by compare the processing of different weighting schemes on various relational and NoSQL database management systems (DBMSs). However, one of the major drawbacks of this previous solution is the fact that is design to work on a single node, as opposed to a distributed environment, which limits its work with large sets of data. In this paper, we seamlessly expand and generalize T$^2$K$^2$D$^2$ to distributed systems to boost its scalability.

For our proposed benchmark solution (\textsc{TextBenDS}), we considered a series of objectives, as follows. We aim to dynamically compute the weighting score at the execution of each new request on the dataset,
thus improving the classical information systems' approach and solving the reproducibility problems~\cite{Lin2016} of computing the weight one time~\cite{Crane2017,Bellot2013}. 
Therefore, when weights are computed, we will take into account the varying size and content changes upon the dataset that happens with time~\cite{Bifet2010}.  

Another objective is to  
redesign the schema to integrate new relationships to store metadata extracted during the preprocessing of the text, i.e., tags and name entities.

Finally, we are aiming at doing extensive experiments
and successfully compare several big data processing solutions. For this paper we will consider the following distributed solutions: Hive - a distributed DBMS~\cite{Thusoo2009}, Spark - a distributed framework~\cite{Zaharia2016} and MongoDB - a document-oriented DBMS. 

As a general criterion,
\textsc{TextBenDS} should be designed with Jim Gray's criteria for a "good" benchmark (relevance, portability, simplicity and scalability)~\cite{Gray1993} in mind.

The remainder of this paper is organized as follows. 
In Section~\ref{sec:sota}, we present a survey on existing big data and, more specifically, text processing-oriented benchmarks, also specifying what our solution brings in addition to the ones presented in this section.
In Section~\ref{sec:specifications}, we describe the general specifications of 
\textsc{TextBenDS},
focusing in particular on the employed data and workload models and on the applied performance metrics.
A detailed description of our distributed implementation can be found in Section~\ref{sec:implementations}. Here we describe our queries, the chosen weighting schemes, as well as the multidimensional implementation for the three selected distributed systems.
In Section~\ref{sec:experiments}, we present the set of experiments for both top-$k$ keywords and documents. We provide an analysis and comparison of the results for each distributed system.
Finally, in Section~\ref{sec:conclusions}, we conclude the paper and provide new research perspectives.

\section{Related Works}\label{sec:sota}

In this section we present an overview of the state of the art related to our contribution.
We briefly analyze and compare some of the most relevant parallel text analysis and processing benchmarks from the big data domain.

Although there are many big data benchmarks that are mostly are data-centric, these solutions focus either on structured data, volume or on MapReduce-based applications, rather than on unstructured or variety. Furthermore, to the best of our knowledge, none deal with directly processing textual data. In these benchmarks, text is used as it is, without further processing or computing different measures, weights, or ranking functions.

For instance, the quasi-standard TPCxHS benchmark models a simple application and features, in addition to classical throughput and response time metrics, availability and energy metrics \cite{tpcxhs15}. BigBench~\cite{Ghazal2013} is the first benchmark that added semi-structured and unstructured data to TPC-DS~\cite{tpcds} and was extended to work on Hadoop and Hive by implementing queries using HiveQL~\cite{Chowdhury2014}. Another improvement to this benchmark, that added additional queries, is BigBench V2~\cite{Ghazal2017}. 
Although, the BigBench benchmark, and its extensions, are developed to work with multidimensional models and semi-structured and unstructured date, their models do not take into account textual data and complex aggregation queries, e.g. queries that compute word weights dynamically or score documents using ranking functions. 

Similarly to BigBench, HiBench~\cite{Huang2010} is a micro-benchmark developed specifically to stress test the capabilities of Hadoop (both MapReduce and HDFS). Using a set of pre-defined Hadoop programs, ranging from data sorting to clustering, HiBench is measuring metrics such as response time, HDFS bandwidth consumption and data access patterns. Another Big Data benchmark is MRBS~\cite{Sangroya2013}. This solution provides workloads of five different domains with the focus on evaluating the dependability of MapReduce systems. 

SparkBench~\cite{Agrawal2016,Li2015} is a micro-benchmark suite developed specifically to stress test the capabilities of Spark on Machine Learning and Graph Computation tasks, rather that text preprocessing and computing weighting schemes. Moreover, Facebook developed LinkBench~\cite{Armstrong2013} to emulate social graph workload on top of databases such as MySQL.

BigDataBench~\cite{Wang2014} features application scenarios from search engines, i.e., the application on Wikipedia entries of operators such as Grep or WordCount. Yet, although BigDataBench is open source, it is quite complex and difficult to extend, especially to test the computation efficiency of term weighting schemes.

BigFUN~\cite{Pirzadeh2015} is a benchmark that uses a synthetic semi-structured social network data in JSON format and it focuses exclusively on micro-operations. The workload consists of queries with various operations such as simple retrieves, range scans, aggregations, joins, as well as inserts and updates. 

By comparison with our solution, all the benchmarks presented above are not used for processing textual data and computing weighting schemes.

There are other types of benchmarks that evaluate parallel text processing in Big Data, cloud applications. However, there are only two available solutions that consist of only specifications without any physical implementation.

The first one is actually a methodology for designing text-oriented benchmarks in Hadoop~\cite{Gattiker2013}. It provides both guidelines and solutions for data preparation and workload definition. Yet, as text analysis benchmarks, its metrics measure the accuracy of analytics results, 
while the objective of our solution is to evaluate the aggregation operations and the computing performance.

The second one is PRIMEBALL~\cite{Ferrarons2014}. It features a fictitious news site hosted in the cloud that is to be managed by the framework under analysis, together with several objective use cases and measures for evaluating system performance . One of its metrics notably involves searching a single word in the corpus. However, PRIMEBALL remains only a specification as of today.

There are also various dedicated text analysis benchmarks that exploit different types of corpora (news articles, movie reviews, books, tweets, synthetic texts...) \cite{OShea2010,Lewis2004,Partalas2015,Kilinc2017,Wang2016}. In terms of metrics, except TextGen \cite{Wang2016} that specifically addresses the performance of word-based compressors, all these benchmarks focus on algorithm accuracy. Either term weights are known before the algorithm is applied, or their computation is incorporated with preprocessing. Furthermore, none of these benchmarks propose adequate data sampling methods based on analysis requirements.

BDGS~\cite{Ming2014} is a benchmark that generates synthetic big data datasets preserving the 4V Big Data properties. BDGS covers three representative data types (structured, semi-structured and unstructured) and three data sources (text, graph, and table data). Although, BDGS generates textual data, the workloads only employ simple computations such as sort, grep and word count. There are other benchmarks that focus on the same workloads~\cite{Jia2014}. The evaluation of complex computation, such as term weighting schemes, are not taken into account by any of these benchmarks. 

FakeNewsNet~\cite{Shu2018} is a text repository of news content, social context and
dynamic information for benchmarking fake news detection, diffusion, and mitigation. The features it presents are linguistic, user profile data, and social network context. The benchmark does not propose any data sampling methods or weighting scheme computation of terms or documents. Furthermore, aggregation methods can be used to combine different features representations into a weighted form and optimize the feature weights by using weighting schemes in the case of textual data~\cite{Shu2017}. 

Thus, the existing text analysis benchmarks do not evaluate weighting schemes construction efficiency. This is why we introduced T$^2$K$^2$~\cite{Truica2017}, a top-$k$ keywords and documents benchmark, and its decision support-oriented evolution T$^2$K$^2$D$^2$~\cite{Truica2018}. Both benchmarks feature a real tweet dataset use case and queries with various complexities and selectivities. They help evaluate weighting schemes and database implementations in terms of computing performance. Yet, these solutions are not tailored for distributed computing. 

As a conclusion to this section, our benchmark (\textsc{TextBenDS}) addresses the following shortcomings that exist in the current literature:
\begin{itemize}
	\item[i)] proposes a number of aggregation queries to compute term weighting schemes and document ranking functions; 
	\item[ii)] tests the computation efficiency of term weighting schemes;
	\item[iii)] offers adequate data sampling methods for analysis;
	\item[iv)] works with structured, semi-structured, and unstructured data in the form of textual data;
	\item[v)] enables analysis based on gender, location, and time to extract general linguistic and social context features. 
\end{itemize}

\section{\textsc{TextBenDS} Specifications}\label{sec:specifications}


In this section, we describe \textsc{TextBenDS}'s data, workload models and  performance metrics; and expand the work done in the original paper~\cite{Truica2018}. This new design creates a generic benchmark that handles not only Twitter datasets, but any kind of textual corpus. 

\subsection{Generic Data Model}\label{sec:T2K2D2data}

For \textsc{TextBenDS}, we remodeled the T$^2$K$^2$D$^2$ multidimensional schema to incorporate new information about the textual documents, thus changing the logical model from a star schema to a snowflake schema. We kept the central fact tables that stores information about the documents, but added new entities to store the tags and named entities.

\textsc{TextBenDS}'s multidimensional snowflake schema is presented in Figure~\ref{fig:ERStarDiagram}.
The models' entities are briefly presented below.

\begin{itemize}
	\item \emph{DocumentFacts} is the central fact entity and contains the number of co-occurrence $f_{t,d}$ and term frequency $TF(t,d)$ for a lemma in a document. 
	The information stored in this table is used for computing the weighting schemes and the ranking functions for extracting the top-$k$ keywords and documents. 
	
	\item \emph{DocumentDimension} is the document dimension table, containing each document's unique identifier and the original and processed text, i.e., original text (\emph{RawText}), clean text (\emph{CleanText}) and lemma text(\emph{LemmaText}). 
	After the top-$k$ documents search process is finished, the information in this entity is used to better visualize the results. 
	
	\item \emph{WordDimension} stores the word's lemma and its unique identifier. The information stored in this entity is a central part of the two tasks at hand.
	It is used to correlate the top ranking terms with the weighting for the the task of top-$k$ keywords. 
	When searching for documents that match specific search queries, a filtering process is applied on the word attribute to select only the documents that contain the search terms before computing the scores for the ranking functions.
	
	\item \emph{TimeDimension} stores the full date and also its hierarchy composed of minute, hour, day, month and year. 
	By filtering this dimension, analysts can apply the roll-up and drill-down operations to better understand the textual data from a time perspective and create different text cubes. Likewise, time series analysis can be applied here, in order to better [understand|interpret] the information. 
	\item \emph{AuthorDimension} stores information about an author's unique identifier, gender, age, firstname, and lastname. By adding constraints on this dimension, analysts can select targeted genders and age ranges for the data mining process.
	\item \emph{LocationDimension} 
	stores the geo-location coordinates for each document. This information is used in filtering to extract only areas of interest for the analysis process.
	
	\item \emph{NamedEntityDimension} stores named entities that extracted after corpus preprocessing. These entities 
	consist of real-world actors, such as
	persons, locations, organizations, products, etc. By adding filtering constraints on this entity, the analysis process can target named entities of interest, thus improving the decision making process offered by business intelligence techniques.
	
	\item \emph{TagDimension} stores the labels for documents, i.e. tags. These tags can be original documents' labels or metadata extracted through preprocessing, such as social media tags. Social media tags include hashtags and mentioning tags, which consists of citing other users’ screen names in tweets (using the syntax \textit{@username}). 
	Hashtags can be used in filtering the dataset based on user interests, while mentioning tags are useful in data mining and graph mining for detecting leaders and followers and to construct the interactive graphs.
	
\end{itemize}

For each dimension, we chose members that best represent 
a generic model of textual data. 
However, the model can be adapted to specific datasets, maybe containing other kinds of textual information.

\begin{figure*}[!htbp]
	\begin{center}
		\includegraphics[width=0.9\columnwidth]{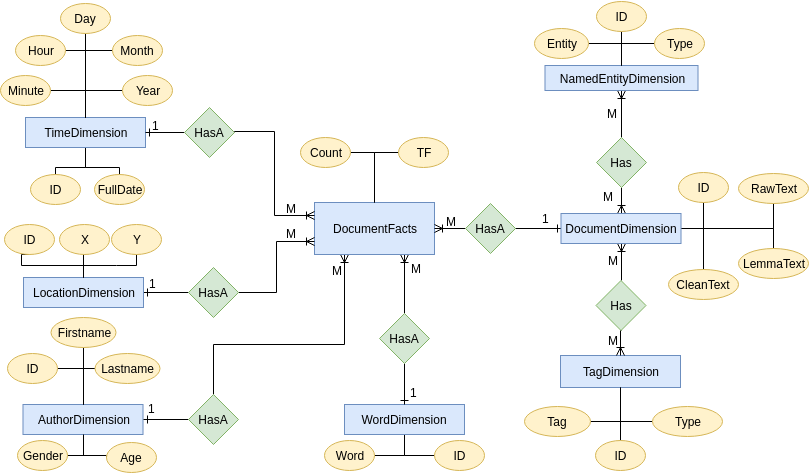}
		\caption{\textsc{TextBenDS} conceptual data model} 
		\label{fig:ERStarDiagram}
	\end{center}
\end{figure*}

\subsection{Workload Model}\label{sec:T2K2D2workload}








The workload model for top-$k$ keywords follows three major analysis directions by subsampling the corpus using filters on gender, time, and location. Also, we further the analysis by extracting top-$k$ documents and subsampling the dataset by adding filters on the \emph{WordDimesion} entity. The workload model uses OLAP queries to extract information from the  multidimensional data model.

We use gender-based filters to extract two sets of features. The first set consist of general linguistic-based features. Using these features, analysts can detect different patterns in writing styles and construct gender-based vocabularies. The second set contains social context features and it's specific to social media datasets. The analysis process based on these features extracts information about gender-based events and objects of interest, as well as psychological profiles and social engagements between authors of the same or of different genders.

We use time-based filters to detect changes that appear over time in the vocabulary, thus extracting topics and events.
The analysis of groups behaviour and the changes it suffers over time can further improve if these filters are associated with the gender-based and location-based filters. Furthermore, when using OLAP operations such as roll-up and drill-down, we can better understand the social context and linguistic based features.

Location-based filters can be used to extract different writing styles and the use of specific vocabularies that may contain regionalisms, archaisms, or idioms. In some cases, these filters ca track events specific to geographic locations, or regional social behaviors such as traditions, fairs, or festivals.

Keyword search filters extract the top-$k$ documents that are similar to the terms in the search query. These filters target specific subjects of interest for the analysis process.
These filters, correlated with the other types of filters, can create a better view for understanding the user opinions regarding products or events, by applying OLAP operations on different dimensions.

The filter categories we discussed above are represented in our model by four constrains, $c_1$ to $c_4$. These constrains are adapted to the models entities and attributes names and are as follows.
\begin{itemize}
	\item $c_1$ is \emph{AutorDimension.Gender = pGender} with  
	\emph{pGender} the gender of the author.
	\item $c_2$ is \emph{TimeDimension.Date $\in$ [pStartDate, pEndDate]}, where 
	\emph{pStartDate $<$ pEndDate} two given dates.
	\item $c_3$ is \emph{LocationDimension.X $\in$ [ pStartX, pEndX]} and \emph{LocationDimension.Y $\in$ [pStartY, pEndY]}, where 
	\emph{pStartX $<$ pEndX}
	and \emph{pStartY $<$ pEndY} 
	are the geo-location coordinates.
	\item $c_4$ is \emph{WordDimension.Word = pTerms} where \emph{pTerms $\in$ \{t $\mid$ t $\in$ vocabulary \} } with \emph{vocabulary} the set of words contained in the corpus.
\end{itemize}


There are many constrains combinations for the filter categories discussed  above.
But, we consider that the most representative constrains for detecting the vocabulary, user behavior, social context, events and opinions
are the following: i) 
for top-k keywords - $c_1$, $c_1 \wedge c_2$, $c_1 \wedge c_3$, and $c_1 \wedge c_2 \wedge c_3$, and ii) for top-$k$ documents -
$c_1 \wedge c_4$, $c_1 \wedge c_2 \wedge c_4$, $c_1 \wedge c_3 \wedge c_4$, and $c_1 \wedge c_2 \wedge c_3 \wedge c_4$.

Eventually, let us emphasize that, in our workload model, filter values are given at execution time  and the scores for extracting the top-$k$ keywords and documents are computed in near-real-time. 
Our model can handle dataset modifications since weights are computed dynamically, therefore removing possible computation errors.
This is an improvement over current information retrieval systems that compute weights only once, when the information is loaded in the database, thus incorporating errors if the ranking the data is modified throught inserts, updates or deletes.



\section{\textsc{TextBenDS} Distributed Implementation}\label{sec:implementations}


In this section, we describe the weighting schemes employed by \textsc{TextBenDS}, as well as the physical multidimensional implementation for Hive and Spark and its translation to the JSON format for MongoDB. 
Moreover, we present \textsc{TextBenDS} queries and discuss their implementation using HiveQL, Spark SQL and Dataframes and the JavaScript MapReduce implementation for MongoDB.

\subsection{Weighting Schemes}
\label{sec:weightingschemas}

A weighting scheme is used in Information Retrieval and Text mining as a statistical measure to evaluate how important a term is to a document in a collection or corpus. The importance increases proportionally to the number of times a term appears in the document but is offset by the frequency of the term in the corpus.

We employ two weighting scheme techniques.
The first  is 
the term frequency-inverse document frequency (TF-IDF) weighting scheme. It is often used as a central tool for scoring and ranking terms in a document.

Given a corpus of documents $D = \{ d_1, d_2, ..., d_N\}$, where $N=|D|$ is the total number of documents in the dataset and $n$ the number of documents where some term $t$ appears. The TF-IDF weight is computed by multiplying the augmented term frequency $TF(t,d) = K + (1 - K) \cdot \frac{f_{t,d}}{\max_{t' \in d}(f_{t',d})}$) by the inverse document frequency $IDF(t,D) = 1 + \log\frac{N}{n}$, i.e., $TFIDF(t,d,D) = TF(t,d) \cdot IDF(t,D)$. The augmented form of $TF$ prevents a bias towards long 
documents when the free parameter $K$ is set to $0.5$~\cite{Paltoglou2010}. It uses the number of co-occurrences $f_{t,d}$ of a word in a document, normalized with the frequency of the most frequent term $t'$, i.e., $\max_{t' \in d}(f_{t',d})$.

The second technique is Okapi BM25, a probabilistic weighting scheme for scoring and ranking documents. It is often used in Information Retrieval and Text mining because it incorporates the document length and the average document length in the corpus to eliminate bias towards long documents.

The Okapi BM25 weight is given in Equation~\eqref{eq:okapi}, where $||d||$ is $d$'s length (i.e., the number of terms appearing in $d$), and  $avg_{d' \in D}(||d'||)$ is the average document length used to remove bias towards long documents. The values of free parameters $k_1$ and $b$ are usually chosen, in absence of advanced optimization, as $k_1 \in [1.2,2.0]$ and $b=0.75$~\cite{Manning2008,SparckJones2000a,SparckJones2000b}.

\begin{equation}\label{eq:okapi}
BM25(t,d,D) = \frac{TFIDF(t,d,D) \cdot (k_1 + 1)}{TF(t,d) + k_1 \cdot (1 - b + b \cdot \frac{||d||}{ avg_{d' \in D}(||d'||)})}
\end{equation}

To extract top-$k$ keywords, the overall relevance of a term $t$ for a given corpus $D$ is computed as the sum of all the TF-IDF (Equation~\eqref{eq:t_TFIDF}) or Okapi BM25 (Equation~\eqref{eq:t_okapi}) weights for that term.

\begin{equation}\label{eq:t_TFIDF}
S_{TK}\_TFIDF(t,D) = \sum_{d_i \in D} TFIDF(t,d_i,D)
\end{equation}

\begin{equation}\label{eq:t_okapi} 
S_{TK}\_BM25(t,D) = \sum_{d_i \in D} BM25(t,d_i,D)
\end{equation}

TF-IDF and Okapi BM25 can be adapted to rank a set of documents based on the search query's terms appearing in each document. Given a search query $Q = \{ q_1, q_2, ..., q_m \}$, where $m=|Q|$ is the number of terms contained in the query, a document $d$ is scored by either summing all the TF-DIF (Equation~\eqref{eq:r_TFIDF}) or the Okapi BM25 (Equation~\eqref{eq:r_okapi}) scores for the query terms in the document. 

\begin{equation}\label{eq:r_TFIDF}
S_{TD}\_TFIDF(Q,d,D) = \sum_{q_i \in Q} TFIDF(q_i,d,D)
\end{equation}

\begin{equation}\label{eq:r_okapi}
S_{TD}\_BM25(Q,d,D) = \sum_{q_i \in Q} BM25(q_i,d,D)
\end{equation}

\subsection{Database Implementation}

\begin{figure*}[!htbp]
	\centering
	\includegraphics[width=0.9\columnwidth]{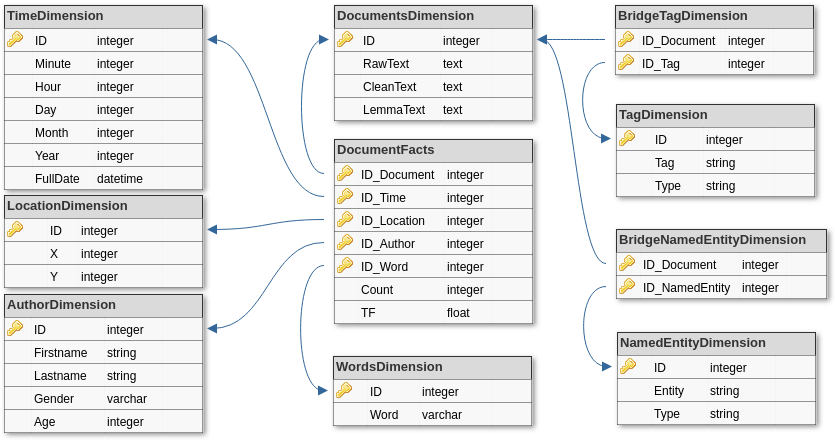}
	\caption{\textsc{TextBenDS} logical snowflake schema} 
	\label{fig:starschema}
\end{figure*}

The conceptual multidimensional snowflake schema described in Section~\ref{sec:specifications} can be directly translated into the database schema presented in Figure~\ref{fig:starschema}. This database schema is used in both Spark framework and Hive data warehouse.

\begin{figure}[!htp]
	\centering
	\includegraphics[width=\columnwidth]{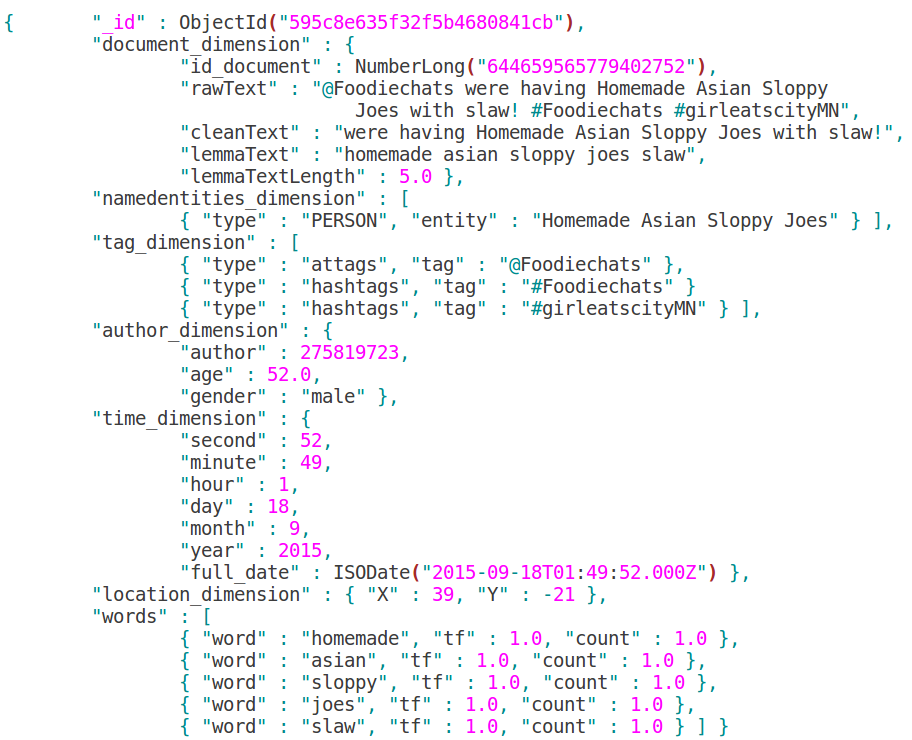}
	\caption{MongoDB document} 
	\label{fig:mongo_doc}
\end{figure}

However, for MongoDB, we need to adjust it to a JSON representation. Thus, when translating the multidimensional snowflake schema into the MongoDB collection, each dimension becomes a nested document inside the same record. Figure~\ref{fig:mongo_doc} presents a document representation.

\subsection{Query Description}

To highlight the performance of the tested distributed data processing and analysis solutions, it is necessary to define a set of complex queries that use aggregation operations and efficiency of computing term weighting schemes. For \textsc{TextBenDS}, we defined two sets of queries. The first set looks for the top-$k$ keywords using the two weighting schemes presented in Equations~\ref{eq:t_TFIDF} and~\ref{eq:t_okapi}, while the second set of queries computes a document ranking scores and extract the top-$k$ documents that are most similar to a search query using the functions based on TF-IDF and Okapi BM25 presented in Equations~\ref{eq:r_TFIDF} and~\ref{eq:r_okapi}.

\subsubsection{Top-$k$ Keywords Queries Set}

In the following paragraphs, we describe the queries from the first set that we used in \textsc{TextBenDS}.

Query $Q_1$ (Equation~\ref{tkk:q1}) computes the top-$k$ keywords for a given gender that is filtered using the constraint $c_1$. This query traverses the \emph{DocumentFacts}, \emph{DocumentDimension}, \emph{WordDimension}, and \emph{AuthorDimension} entities and uses two JOIN constraints between them:
\textit{i)} $c_5$ is the JOIN constraint between the \emph{DocumentFacts} and the \emph{WordDimension} entities and
\textit{ii)} $c_6$ is the JOIN constraint between the \emph{DocumentFacts} and the \emph{AuthorDimension} entities.
To limit the query to extract only the top-$k$ results we use the constraint $c_{tk}$. This query is the basis for all the other top-$k$ keywords queries.

\begin{multline}~\label{tkk:q1}
	Q_1 = \sigma_{c_{tk}}( \gamma_L( \pi_{WordDimension.Word, f_w}( \sigma_{c_1}( DocumentFacts \bowtie_{c_5} \\ WordDimension  \bowtie_{c_6} AuthorDimension ))))
\end{multline}

Query $Q_2$ (Equation~\ref{tkk:q2}) is a direct modification of $Q_1$ that filters the results by gender for a given time window. Besides the \emph{DocumentFacts}, \emph{DocumentDimension}, \emph{WordDimension}, and \emph{AuthorDimension} entities, $Q_2$ also traverses the \emph{TimeDimension} by adding the JOIN constraint $c_7$ between \emph{DocumentFacts} and \emph{TimeDimension} entities. The filtering constraint on the \emph{TimeDimension} is $c_2$.

\begin{multline}~\label{tkk:q2}
	Q_2 = \sigma_{c_{tk}}( \gamma_L( \pi_{WordDimension.Word, f_w}( \sigma_{c_1 \wedge c_2}( DocumentFacts \bowtie_{c_5} \\ WordDimension \bowtie_{c_6} AuthorDimension \bowtie_{c_7} TimeDimension ))))
\end{multline}      

Query $Q_3$ (Equation~\ref{tkk:q3}) is also a direct modification of $Q_1$, but this time the results are filtered by gender for a given geographic area. $Q_2$ traverses the \emph{LocationDimension} by adding the JOIN constraint $c_8$ between \emph{DocumentFacts} and \emph{LocationDimension} entities. The other three entities from $Q_1$ with their respective JOIN constraint remain the same. For filtering the results for a given geographic area, the query uses the constraint $c_3$.

\begin{multline}~\label{tkk:q3}
	Q_3 = \sigma_{c_{tk}}( \gamma_L( \pi_{WordDimension.Word, f_w}( \sigma_{c_1 \wedge c_3}( DocumentFacts \bowtie_{c_5} \\ WordDimension  \bowtie_{c_6} AuthorDimension \bowtie_{c_8} LocationDimension ))))
\end{multline}      

Finally, $Q_4$ (Equation~\ref{tkk:q4}) is used to determine the top-$k$ keywords for authors who have a given gender for a time window and a geographic area. This query traverses the \emph{DocumentFacts}, \emph{DocumentDimension}, \emph{WordDimension}, \emph{AuthorDimension}, \emph{TimeDimension}, and \emph{LocationDimension} entities. The constrains $c_5$ to $c_8$ are uses to JOIN the \emph{DocumentFacts} with each dimension entity. Constraints $c_1$ to $c_3$ are employed to filter the results for a given gender, a time window given by two dates, and a geographic area given by the geo-location coordinates. 

\begin{multline}~\label{tkk:q4}
	Q_4 = \sigma_{c_{tk}}( \gamma_L( \pi_{WordDimension.Word, f_w}( \sigma_{c_1 \wedge c_2 \wedge c_3}( DocumentFacts \bowtie_{c_5} \\ WordDimension \bowtie_{c_6} AuthorDimension \bowtie_{c_7} TimeDimension \bowtie_{c_8} \\ LocationDimension ))))
\end{multline}

In all four queries, $f_w$ is used to compute the weighting scheme using nested aggregation queries. To compute TF-IDF, we need to determine: \textit{i)} the term frequency, \textit{ii)} $N$ the total number of documents in the corpus, and \textit{iii)} $n$ the number of documents where a term appears. The term-frequency is already known, as it is present in the \emph{DocumentFacts} and can be extracted as follows $TF(t,d) = DocumentFacts.TF$. A nested query with the count aggregation function is used to compute $N$. This query takes into account the filters present in each of the four queries. To compute $n$, we just count how many times each a word appears given the constraints of each query $Q_1$ to $Q_4$.  

For Okapi BM25, besides what we compute for the TF-IDF weighting, i.e. $n$, $N$, and $TF(t,d)$, we also need the length of each document $DL$. To calculate the $DL$, a nested query is used to filter the data using the constrains of each query $Q_1$ to $Q_4$. Then we use the aggregation function $sum$ with the GROUP BY clause on the document's unique identifier. This function computes the document length by adding the number of appearances of each word in the document. 

To determine the weights of each word, regardless of the weighting scheme, we use the aggregation operator $\gamma_L$, where $L=(F, G)$. The list of aggregation functions is given by $L$, while the set of attributes in the GROUP BY clause is given by $G$. The list of aggregation functions is $F = \{ sum(f_w) \}$, where the $sum$ is the aggregation function that computes $S_{TK}\_TFIDF(t,D)$ (Equation~\eqref{eq:t_TFIDF}). The set of attributes in the GROUP BY clause is $G = \{ WordDimension.Word \}$.

\subsubsection{Top-$k$ Documents Queries Set}

In top-$k$ documents queries (in addition to top-$k$ keywords queries), we add constraint $c_4$  to select the documents that contain the search terms in list $Q$. 

Top-$k$ documents query $Q'_1$ (Equation~\ref{tkd:q1}) uses the same filters as top-$k$ keywords $Q_1$. 
In addition, we add another filter to retrieve only the documents matching a given search query $Q$. 
The search query filtering constraint is $c_4$. Thus, query $Q'_1$ determines the top-$k$ documents for authors who have a given gender for a search query $Q$.

\begin{multline}~\label{tkd:q1}
	Q'_1 = \sigma_{c_{tk}}( \gamma_L( \pi_{DocumentFacts.ID\_Document, f_w}( \sigma_{c_1 \wedge c_4}( DocumentFacts \bowtie_{c_5} \\ WordDimension \bowtie_{c_6} AuthorDimension ))))
\end{multline}      

Query $Q'_2$ (Equation~\ref{tkd:q2}) determines the top-$k$ documents for a search query after filtering the records using constraints on the gender and a time window. 

\begin{multline}~\label{tkd:q2}
	Q'_2 = \sigma_{c_{tk}}( \gamma_L( \pi_{DocumentFacts.ID\_Document, f_w}( \sigma_{c_1 \wedge c_2 \wedge c_4}( DocumentFacts \bowtie_{c_5} \\ WordDimension \bowtie_{c_6} AuthorDimension \bowtie_{c_7} TimeDimension ))))
\end{multline}      

Query $Q'_3$ (Equation~\ref{tkd:q2}) filters the results by the authors' gender and the geographic area, after using the terms in the search query to extract the relevant top-$k$ documents.

\begin{multline}~\label{tkd:q3}
	Q'_3 = \sigma_{c_{tk}}( \gamma_L( \pi_{DocumentFacts.ID\_Document, f_w}( \sigma_{c_1 \wedge c_3 \wedge c_4}( DocumentFacts \bowtie_{c_5} \\ WordDimension \bowtie_{c_6} AuthorDimension \bowtie_{c_8} LocationDimension  ))))
\end{multline}      

Finally, as in the case of the last top-$k$ keywords query, query $Q_4$ (Equation~\ref{tkd:q4}) filters the results by the author's gender, a time window, and a geographic area and then extracts the top-$k$ documents relevant to the search query.

\begin{multline}~\label{tkd:q4}
	Q'_4 = \sigma_{c_{tk}}( \gamma_L( \pi_{DocumentFacts.ID\_Document, f_w}( \sigma_{c_1 \wedge c_2 \wedge c_3 \wedge c_4}( DocumentFacts \bowtie_{c_5} \\ WordDimension \bowtie_{c_6} AuthorDimension \bowtie_{c_7} TimeDimension \bowtie_{c_8} \\ LocationDimension ))))
\end{multline}


The $f_w$ function is computed exactly as in the case of the top-$k$ keywords queries, i.e. using nested aggregation queries. But, the aggregation operator $\gamma_L$ is different from the one in the top-$k$ keywords, i.e., the grouping is done using the document unique identifier ($G = \{ DocumentFacts.ID\_Document \}$) and the list of aggregation function $F = \{ sum(f_w) \}$.
To compute the hierarchy of documents with TF-IDF, we use the following formula $sum(f_w) = S_{TD}\_TFIDF(Q,d,D)$ (Equation~\eqref{eq:r_TFIDF}), while for Okapi BM25  $sum(f_w) = S_{TD}\_BM25(Q,d,D)$ (Equation~\eqref{eq:r_okapi}). 
$Q$ represents the list of search terms, in both functions described above.

For Hive, we have implemented the queries using HiveQL (Hive Query Language) a SQL-live query language that uses implicitly MapReduce or Tez~\cite{Saha2015}. For Spark, the queries were implemented using Spark SQL together with Spark Dataframes~\cite{Armbrust2015}. For MongoDB, we implemented the queries using the native JavaScript language API using the MapReduce (MR) framework for all the queries. 
In the case of the top-$k$ keywords queries that use the TF-IDF weighting scheme, we take advantage of the native database aggregation framework, i.e., the aggregation pipeline (AP).

\section{Experiments}\label{sec:experiments}

In this section, 
we first present \textsc{TextBenDS}'s performance metrics and execution protocol.
Second, we present the hardware architecture and the software configuration used for our benchmarking experiment. 
Third, we discuss the dataset, as well as the query complexity and selectivity. 
Forth, we present a comparison by data management distributed system and weighting scheme. 
Finally, we compare the runtime of these data management distributed system implementations w.r.t. the scale factor ($SF$) and weighting schemes.

\subsection{Performance Metrics and Execution Protocol}
\label{sec:MetricsProtocol}

For \textsc{TextBenDS}, we use as metric only the query response time.
We note the response time for each query as $t(Q_i )$ and $t(Q'_i )$ $\forall i \in [1, 4]$.  All queries $Q_1$ to $Q_4$ and $Q'_1$ to $Q'_4$ are executed $10$ times for both top-$k$ keywords and top-$k$ documents, which is sufficient according to the central limit theorem. Average response times and standard deviations are computed for $t(Q_i)$ and $t(Q'_i)$. All executions are warm runs, i.e., either caching mechanisms must be deactivated, or a cold run of $Q_1$ to $Q_4$ and $Q'_1$ to $Q'_4$ must be executed once (but not taken into account in the benchmark's results) to fill in the cache. Queries must be written in the native scripting language of the target database system and executed directly inside said system using the command line interpreter. 

\subsection{Experimental conditions}

All tests run on a cluster with 6 nodes running Ubuntu 16.04 x64, each with 1 Intel Core i7-4790S CPU with 8 cores at 3.20GHz, 16 GB RAM and 500GB HDD. The Hadoop ecosystem is running on Ambari and has the following configuration for the 6 nodes: 1 node acts as HDFS Name Node and Secondary Name Node, YARN Resource Manager, YARN Application Manager, the Hive Services and the Spark Driver and 5 nodes, each acting as HDFS Data Nodes and YARN Node Managers. YARN, Hive, Spark, Tez, and MapReduce Clients are installed on all the nodes.

The HDFS~\cite{Shvachko2010} Name Node is a master node that stores the directory tree of the file system, file metadata, and the locations of each file in the cluster. The Secondary Name Node is a master node that performs housekeeping tasks and checkpointing on behalf of the Name Node. The HDFS Data Node is a worker node that stores and manages HDFS blocks on the local disk. 

The YARN Resource Manager~\cite{Vavilapalli2013} is a master node that allocates and monitors available cluster resources (e.g., physical assets like memory and processor cores) to applications as well as handling scheduling of jobs on the cluster. The YARN Resource Manager is a master node that coordinates a particular application being run on the cluster as scheduled by the Resource Manager. The YARN Node Manager is a worker node that runs and manages processing tasks on an individual node as well as reports the health and status of tasks as they are running. 

The Hive Services deal with the client interactions with Hive. The main component of Hive Sevices is the Hive Driver which processes all the requests from different applications to the Hive Metastore and The Hive File System for further processing. The Hive Metastore is the central repository of Apache Hive metadata. The Hive File System communicates with the Hive Storage which usually is built on top of HDFS. 

The Hive Clients provide different drivers for communication with different types of applications. We use Tez as the Hive query execution engine. Tez~\cite{Saha2015} is an extensible framework for building high performance batch and interactive data processing applications coordinated by YARN that improves the MapReduce~\cite{Dean2008} paradigm by dramatically enhancing its speed, while maintaining MapReduce's ability to scale to petabytes of data. 

The Spark Driver is the master node for the application and uses the Task Scheduler to launches task via cluster manager, i.e. YARN, and Directed Acyclic Graph (DAG) Scheduler used to divide operators into stages of tasks. The Spark Clients are the worker nodes.

For the MongoDB tests, we used the same cluster infrastructure with the following specifications: 1 node with the MongoDB Configuration Server and MongoDB Shard Server (\textit{mongos}) and 5 nodes with MongoDB shards. The MongoDB Configuration Server stores the metadata for a sharded cluster. This metadata reflects state and organization for all data and components within the sharded cluster. The MongoDB Shard Server is a routing service for the MongoDB shard configurations. It processes queries from the application layer, and determines the location of this data in the sharded cluster, in order to complete these operations.



The number of Spark executors was fixed to $16$ with one vnode and 3GB memory each for the Spark experiments. Moreover, we use the Spark SQL and Dataframes libraries for the Spark experiments together with the Scala programming language. The Hive Server Heap Size and the Hive Metastore Heap Size are both set to 2GB each, while the Hive Client Heap Size is set to 1G. Each reducer can process 1GB of data at a time. The dataset is stored on HDFS for both Hive and Spark experiments under the ORC format. 

The code of all Hive queries and Scala code for the Spark experiments, together with benchmarking results, are available on Github\footnote{Source code~\url{https://github.com/cipriantruica/T2K2D2_Hadoop}}

The query parameterization is provided in Table~\ref{tbl:qparam}.

\begin{table}[!ht]
	\centering
	\caption{Query parameter values}
	\label{tbl:qparam}
	\begin{tabular}{|l|c|}
		\hline
		\textbf{Parameter}  & \textbf{Value}             \\ \hline
		\textit{pGender}    & $\{male, female\}$         \\ \hline
		\textit{pStartDate} & 2015-09-17 00:00:00        \\ \hline
		\textit{peEndDate}  & 2015-09-18 00:00:00        \\ \hline
		\textit{pStartX}    & 20                         \\ \hline
		\textit{pEndX}      & 40                         \\ \hline
		\textit{pStartY}    & -100                       \\ \hline
		\textit{pEndY}      & 100                        \\ \hline
		\textit{pWords}     & $\{think, today, friday\}$ \\ \hline
	\end{tabular}
\end{table}

\subsection{Dataset}

The experiments are done on a 2\,500\,000 tweets corpus. The initial corpus is split into $5$ different datasets equally balanced between the number of tweets for gender, location, and date. These datasets contain 500\,000, 1\,000\,000, 1\,500\,000, 2\,000\,000, and 2\,500\,000 tweets, respectively. They allow scaling experiments and are associated to a scale factor ($SF$) parameter, where $SF \in \{0.5, 1, 1.5, 2, 2.5\}$.

\subsubsection{Query selectivity}

Selectivity, i.e., the amount of retrieved data ($n(Q)$) w.r.t. the total amount of data available ($N$), depends on the number of attributes in the WHERE and GROUP BY clauses. The selectivity formula used for a query $Q$ is $S(Q) = 1-\frac{n(Q)}{N}$.

\textsc{TextBenDS}'s queries traverse the \emph{DocumentFacts}, \emph{WordDimnesion}, and \emph{AuthorDimension} relationships.

All queries filter by gender, to determine the trending words for female (F) and male (M) users. Starting from $Q_1$, subsequent queries $Q_2$ to $Q_4$ are built by decreasing selectivity (Table~\ref{tbl:selectivity}). Moreover, by adding a constraint on the location in $Q_3$ and $Q_4$, the query complexity also changes.

\begin{table}[!ht]
	\centering
	\caption{Top-$k$ keywords query selectivity}
	\label{tbl:selectivity}
	\begin{tabular}{|c|c|c|c|c|c|c|c|c|}
		\hline
		\textbf{\textit{SF}} & \textbf{\textit{$Q_1$ (M)}} & \textbf{\textit{$Q_1$ (F)}} & \textbf{\textit{$Q_2$ (M)}} & \textbf{\textit{$Q_2$ (F)}} & \textbf{\textit{$Q_3$ (M)}} & \textbf{\textit{$Q_3$ (F)}} & \textbf{\textit{$Q_4$ (M)}} & \textbf{\textit{$Q_4$ (F)}} \\ \hline
		0.5 & 0.336 & 0.337 & 0.517 & 0.517 & 0.556 & 0.558 & 0.677 & 0.679  \\ \hline
		1   & 0.342 & 0.342 & 0.662 & 0.662 & 0.562 & 0.565 & 0.774 & 0.775  \\ \hline
		1.5 & 0.347 & 0.346 & 0.736 & 0.736 & 0.569 & 0.572 & 0.823 & 0.824  \\ \hline
		2   & 0.351 & 0.350 & 0.783 & 0.783 & 0.574 & 0.575 & 0.855 & 0.856  \\ \hline
		2.5 & 0.353 & 0.354 & 0.815 & 0.815 & 0.579 & 0.580 & 0.876 & 0.877  \\ \hline
	\end{tabular}
\end{table}

Table~\ref{tbl:selectivity_docs_3w} presents the  selectivity for the top-$k$ documents.
Compared to top-$k$ keywords, the selectivity for queries $Q'_1$ to $Q'_4$ decreases even more by adding a condition on the words attribute for all the queries.

\begin{table}[!ht]
	\centering
	\caption{Top-$k$ documents selectivity for 3 search terms}
	\label{tbl:selectivity_docs_3w}
	\begin{tabular}{|c|c|c|c|c|c|c|c|c|}
		\hline
		\textbf{\textit{SF}} & \textbf{\textit{$Q'_1$ (M)}} & \textbf{\textit{$Q'_1$ (F)}} & \textbf{\textit{$Q'_2$ (M)}} & \textbf{\textit{$Q'_2$ (F)}} & \textbf{\textit{$Q'_3$ (M)}} & \textbf{\textit{$Q'_3$ (F)}} & \textbf{\textit{$Q'_4$ (M)}} & \textbf{\textit{$Q'_4$ (F)}} \\ \hline
		0.5 & 0.9844 & 0.9848 & 0.9904 & 0.9905 & 0.9921 & 0.9926 & 0.9951 & 0.9954 \\ \hline
		1   & 0.9866 & 0.9868 & 0.9952 & 0.9953 & 0.9932 & 0.9936 & 0.9975 & 0.9977 \\ \hline
		1.5 & 0.9835 & 0.9837 & 0.9968 & 0.9968 & 0.9917 & 0.9920 & 0.9984 & 0.9985 \\ \hline
		2   & 0.9822 & 0.9824 & 0.9976 & 0.9976 & 0.9910 & 0.9913 & 0.9988 & 0.9988 \\ \hline
		2.5 & 0.9825 & 0.9827 & 0.9981 & 0.9981 & 0.9912 & 0.9915 & 0.9990 & 0.9991 \\ \hline
	\end{tabular}
\end{table}

\subsubsection{Query complexity}

Complexity relates to the number of traversals involved in the query. Query complexity depends on the number of relationship and entity traversals. Independently from any weighting scheme, all the queries traverse the  \emph{DocumentFacts}, \textit{DocumentDimension}, \textit{WordDimension}, \textit{AuthorDimension} in order to determine the top-$k$ keywords and documents. Regardless of the query, we will call these traversals the "main part" of the query $Q_M$.

To compute the top-$k$ keywords using TF-IDF, we need a new query, $Q_{nD}$, that determines the number of documents. This query is used in the projection section of each query. 
The base of $Q_{nD}$ traverses \emph{DocumentFacts} and \textit{AuthorDimension} and is used in the projection section of $Q_M$.
To compute the top-$k$ keywords using Okapi BM25, another query is added, $Q_{DL}$, which determines the document length. This query traverses the \textit{DocumentFacts} and \textit{AuthorDimension} and it needs to be traversed in $Q_M$.
Table~\ref{tbl:tkk_traverse} presents the dimensions traversed and the nested queries required to compute the top-$k$ keywords.

\begin{table}[!ht]
	\centering
	\caption{Top-$k$ keywords dimension traversals}
	\label{tbl:tkk_traverse}
	\begin{tabular}{|l|l|l|c|c|}
		\hline
		\multicolumn{1}{|c|}{\multirow{2}{*}{\textbf{Query}}} & \multicolumn{1}{c|}{\multirow{2}{*}{\textbf{\begin{tabular}[c]{@{}c@{}}Weighting\\ Scheme\end{tabular}}}} & \multicolumn{1}{c|}{\multirow{2}{*}{\textbf{\begin{tabular}[c]{@{}c@{}}Nested\\ Queries\end{tabular}}}} & \multicolumn{2}{c|}{\textbf{Traversed Relationships}} \\ \cline{4-5} 
		\multicolumn{1}{|c|}{} & \multicolumn{1}{c|}{} & \multicolumn{1}{c|}{} & \textbf{\textit{TimeDimension}} & \textbf{\textit{LocationDimension}} \\ \hline
		\multirow{3}{*}{$Q_1$} & Both & $Q_M$ & $\times$ & $\times$ \\ \cline{2-5} 
		& TF-IDF & $Q_{nD}$ & $\times$ & $\times$ \\ \cline{2-5} 
		& Okapi BM25 & $Q_{DL}$ & $\times$ & $\times$ \\ \hline 
		\multirow{3}{*}{$Q_2$} & Both & $Q_M$ & $\checkmark$ & $\times$ \\ \cline{2-5} 
		& TF-IDF & $Q_{nD}$ & $\checkmark$ & $\times$ \\ \cline{2-5} 
		& Okapi BM25 & $Q_{DL}$ & $\checkmark$ & $\times$ \\ \hline
		\multirow{3}{*}{$Q_3$} & Both & $Q_M$ & $\times$ & $\checkmark$ \\ \cline{2-5} 
		& TF-IDF & $Q_{nD}$ & $\times$ & $\checkmark$ \\ \cline{2-5} 
		& Okapi BM25 & $Q_{DL}$ & $\times$ & $\checkmark$ \\ \hline
		\multirow{3}{*}{$Q_4$} & Both & $Q_M$ & $\checkmark$ & $\checkmark$ \\ \cline{2-5} 
		& TF-IDF & $Q_{nD}$ & $\checkmark$ & $\checkmark$ \\ \cline{2-5} 
		& Okapi BM25 & $Q_{DL}$ & $\checkmark$ & $\checkmark$ \\ \hline
	\end{tabular}
\end{table}

Table~\ref{tbl:query_complexity_star} present the queries complexity for the top-$k$ keywords. The complexity of each query adds to that of $Q_1$ the number of traversals presented in Table~\ref{tbl:tkk_traverse}. As expected, $Q_4$ has the highest complexity. 
As $Q_DL$ is also traversed in $Q_M$ when using Okapi BM25, there is a difference of 1 between the two weighting schemes.

\begin{table}[!ht]
	\centering
	\caption{Query complexity for top-$k$ keywords}
	\label{tbl:query_complexity_star}
	\begin{tabular}{l|c|c|c|c|}
		\cline{2-5}
		& \textbf{\textit{$Q_1$}} & \textbf{\textit{$Q_2$}} & \textbf{\textit{$Q_3$}} & \textbf{\textit{$Q_4$}} \\ \hline
		\multicolumn{1}{|l|}{\textbf{TF-IDF}}      & 3          & 5          & 5          & 7          \\ \hline
		\multicolumn{1}{|l|}{\textbf{Okapi BM25}} & 4          & 6          & 6          & 8          \\ \hline
	\end{tabular}
\end{table}

To compute the top-$k$ documents, in addition to the top-$k$ keywords queries, $Q_{nD}$ for TF-IDF and $Q_{DL}$ for Okapi BM25, we use the $Q_{nW}$ query that computes the number of words for each document. This query is needed for both weighting schemes and traverses the \textit{DocumentFacts} and \textit{AuthorDimension}. The relationship obtained by $Q_{nW}$ query is traversed in the main part of each $Q'_M$ query. 
$Q_{nD}$ query is used in the projection section of each query in order to determine the top-$k$ documents using TF-IDF, whereas if we want to determine the top-$k$ documents using Okapi BM25, $Q_{DL}$ query needs to be traversed in $Q'_M$.
Table~\ref{tbl:tkd_traverse} presents the dimensions traversed and the nested queries required to compute the top-$k$ documents.


\begin{table}[!ht]
	\centering
	\caption{Top-$k$ documents dimension traversals}
	\label{tbl:tkd_traverse}
	\begin{tabular}{|l|l|l|c|c|}
		\hline
		\multicolumn{1}{|c|}{\multirow{2}{*}{\textbf{Query}}} & \multicolumn{1}{c|}{\multirow{2}{*}{\textbf{\begin{tabular}[c]{@{}c@{}}Weighting\\ Scheme\end{tabular}}}} & \multicolumn{1}{c|}{\multirow{2}{*}{\textbf{\begin{tabular}[c]{@{}c@{}}Nested\\ Queries\end{tabular}}}} & \multicolumn{2}{c|}{\textbf{Traversed Relationships}} \\ \cline{4-5} 
		\multicolumn{1}{|c|}{} & \multicolumn{1}{c|}{} & \multicolumn{1}{c|}{} & \textbf{\textit{TimeDimension}} & \textbf{\textit{LocationDimension}} \\ \hline
		\multirow{4}{*}{$Q'_1$} & Both & $Q'_M$ & $\times$ & $\times$ \\ \cline{2-5} 
		& Both & $Q_{nW}$ & $\times$ & $\times$ \\ \cline{2-5} 
		& TF-IDF & $Q_{nD}$ & $\times$ & $\times$ \\ \cline{2-5} 
		& Okapi BM25 & $Q_{DL}$ & $\times$ & $\times$ \\ \hline
		\multirow{4}{*}{$Q'_2$} & Both & $Q'_M$ & $\checkmark$ & $\times$ \\ \cline{2-5} 
		& Both & $Q_{nW}$ & $\checkmark$ & $\times$ \\ \cline{2-5} 
		& TF-IDF & $Q_{nD}$ & $\checkmark$ & $\times$ \\ \cline{2-5} 
		& Okapi BM25 & $Q_{DL}$ & $\checkmark$ & $\times$ \\ \hline
		\multirow{4}{*}{$Q'_3$} & Both & $Q'_M$ & $\times$ & $\checkmark$ \\ \cline{2-5} 
		& Both & $Q_{nW}$ & $\times$ & $\checkmark$ \\ \cline{2-5} 
		& TF-IDF & $Q_{nD}$ & $\times$ & $\checkmark$ \\ \cline{2-5} 
		& Okapi BM25 & $Q_{DL}$ & $\times$ & $\checkmark$ \\ \hline
		\multirow{4}{*}{$Q'_4$} & Both & $Q'_M$ & $\checkmark$ & $\checkmark$ \\ \cline{2-5} 
		& Both & $Q_{nW}$ & $\checkmark$ & $\checkmark$ \\ \cline{2-5} 
		& TF-IDF & $Q_{nD}$ & $\checkmark$ & $\checkmark$ \\ \cline{2-5} 
		& Okapi BM25 & $Q_{DL}$ & $\checkmark$ & $\checkmark$ \\ \hline
	\end{tabular}
\end{table}

When using TF-IDF with $Q'_1$ query, only the results obtained by $Q_{nW}$ need to also be traversed in $Q'_M$. $Q_DL$, beside computing the length of each document, is also used to count the number of documents in the projection section of $Q'_M$.

For the Okapi BM25, both $Q_{nW}$ and $Q_DL$ are traversed in $Q'_M$, thus increasing the complexity by $1$. 
For $Q'_2$ and $Q'_3$ the \textit{TimeDimension} and \textit{LocationDimension} relationships need to be traversed in $Q'_M$ , $Q_{nW}$, and $Q_DL$, thus increasing the complexity of $Q'_1$ by $3$.
For $Q'_4$ both the \textit{TimeDimension} and \textit{LocationDimension} relationships need to be traversed in $Q_M$, $Q_{nW}$, and $Q_DL$, thus increasing the complexity of $Q'_1$ by $6$.

The queries complexity for top-$k$ documents is presented in Table~\ref{tbl:query_complexity_star_docs}.

\begin{table}[!htb]
	\centering
	\caption{Query complexity for top-$k$ documents}
	\label{tbl:query_complexity_star_docs}
	\begin{tabular}{l|c|c|c|c|}
		\cline{2-5}
		& \textbf{\textit{$Q'_1$}} & \textbf{\textit{$Q'_2$}} & \textbf{\textit{$Q'_3$}} & \textbf{\textit{$Q'_4$}} \\ \hline
		\multicolumn{1}{|l|}{\textbf{TF-IDF}}      & 5        & 8        & 8       & 11          \\ \hline
		\multicolumn{1}{|l|}{\textbf{Okapi BM25}} & 6        & 9        & 9       & 12          \\ \hline
	\end{tabular}
\end{table}

\subsection{Weighting Scheme Comparison}

\begin{figure*}[!htbp]
	\centering
	\begin{subfigure}{0.9\columnwidth}
		\centering
		\includegraphics[width=\columnwidth]{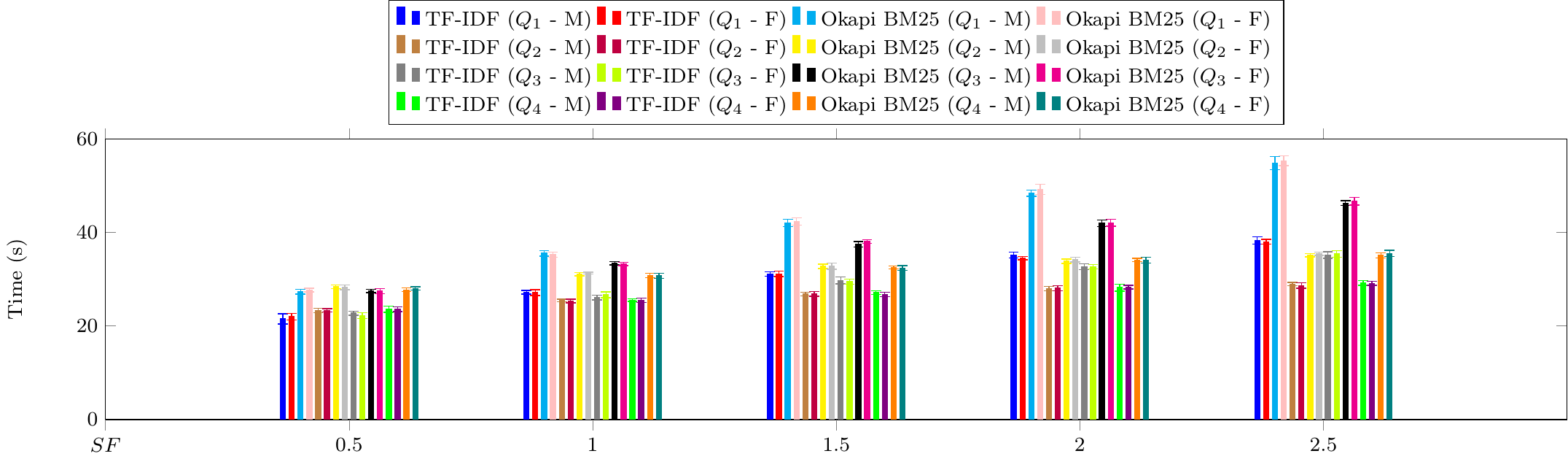}
		\caption{Hive top-$k$ keywords: TF-IDF vs. Okapi BM25}
		\label{fig:hive_tkk}
	\end{subfigure}
	\begin{subfigure}{0.9\columnwidth}
		\centering
		\includegraphics[width=\columnwidth]{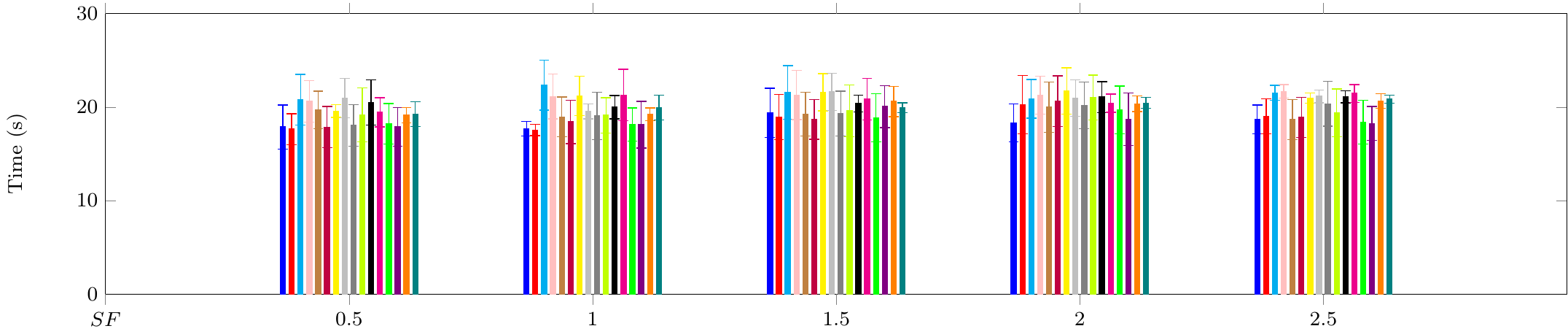}
		\caption{Spark top-$k$ keywords: TF-IDF vs. Okapi BM25}
		\label{fig:spark_tkk}
	\end{subfigure}
	\begin{subfigure}{0.9\columnwidth}
		\centering
		\includegraphics[width=\columnwidth]{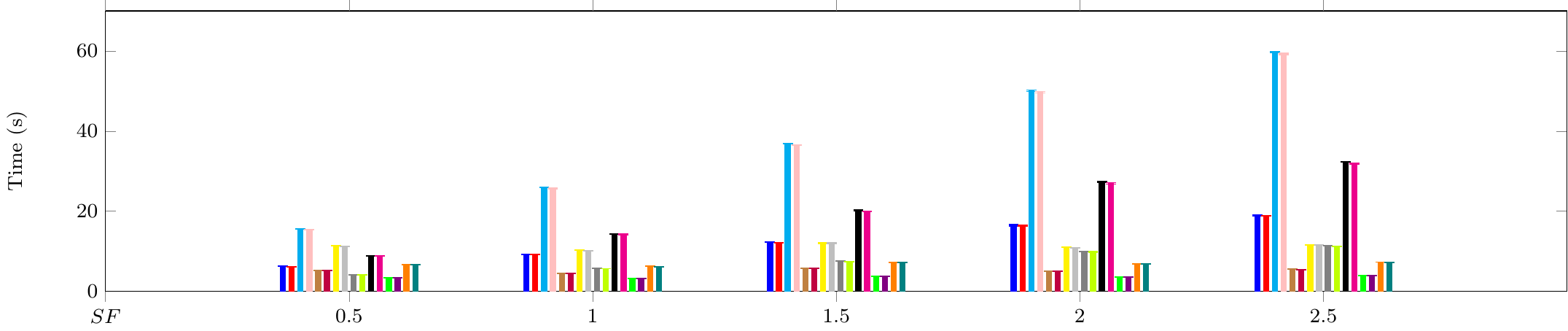}
		\caption{MongoDB MR top-$k$ keywords: TF-IDF vs. Okapi BM25}
		\label{fig:mongo_tkk}
	\end{subfigure}
	\caption{Top-$k$ keywords: TF-IDF vs. Okapi BM25 comparison}
	\label{fig:comp_tkk_tfidf_okapi}
\end{figure*}
\begin{figure*}[!htbp]
	\centering
	\begin{subfigure}{0.9\columnwidth}
		\centering
		\includegraphics[width=\columnwidth]{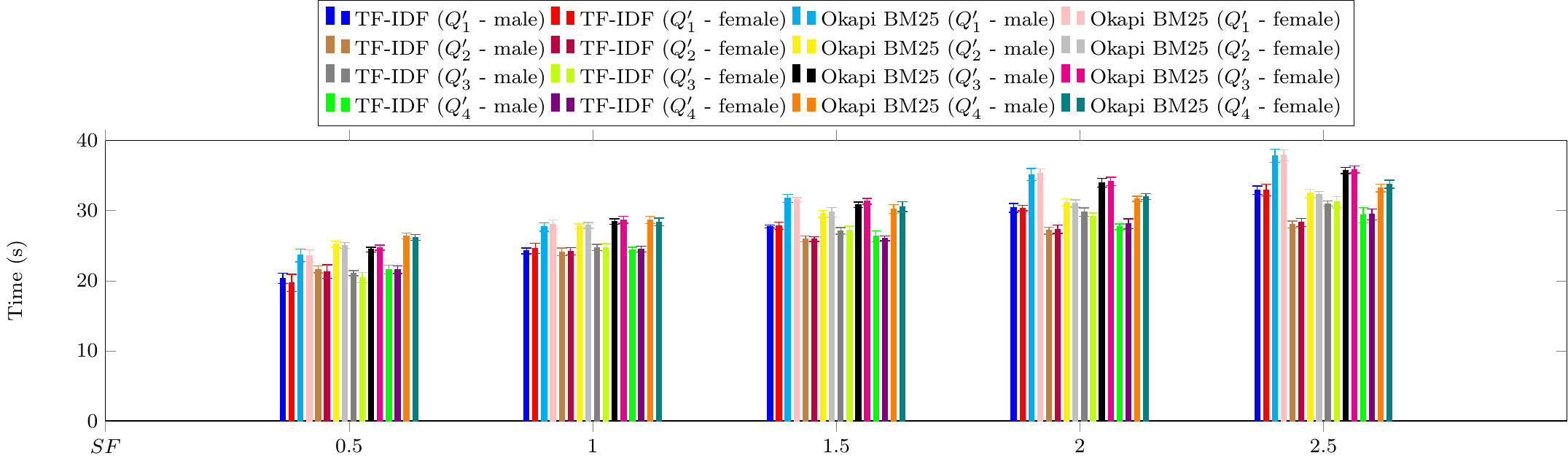}
		\caption{Hive top-$k$ documents: TF-IDF vs. Okapi BM25}
		\label{fig:hive_tkd}
	\end{subfigure}
	\begin{subfigure}{0.9\columnwidth}
		\centering
		\includegraphics[width=\columnwidth]{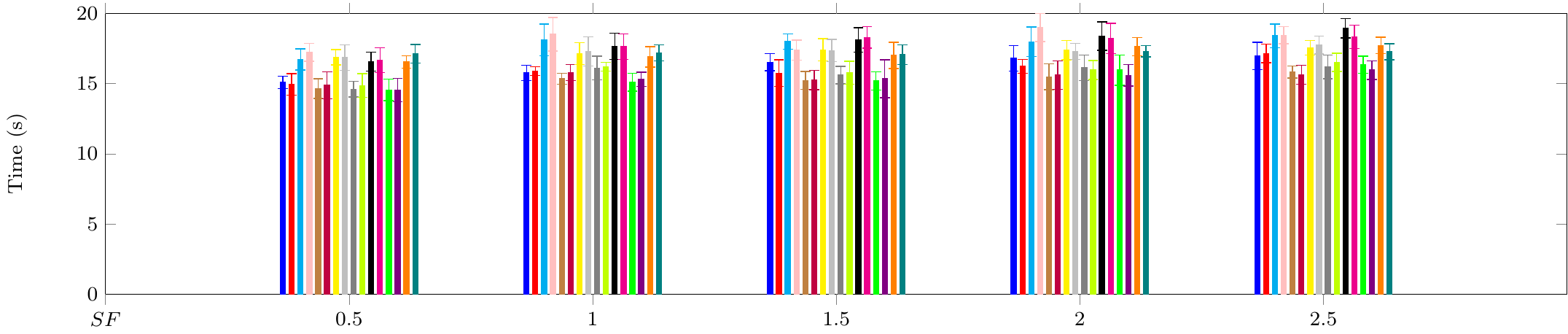}
		\caption{Spark top-$k$ documents: TF-IDF vs. Okapi BM25}
		\label{fig:spark_tkd}
	\end{subfigure}
	\begin{subfigure}{0.9\columnwidth}
		\centering
		\includegraphics[width=\columnwidth]{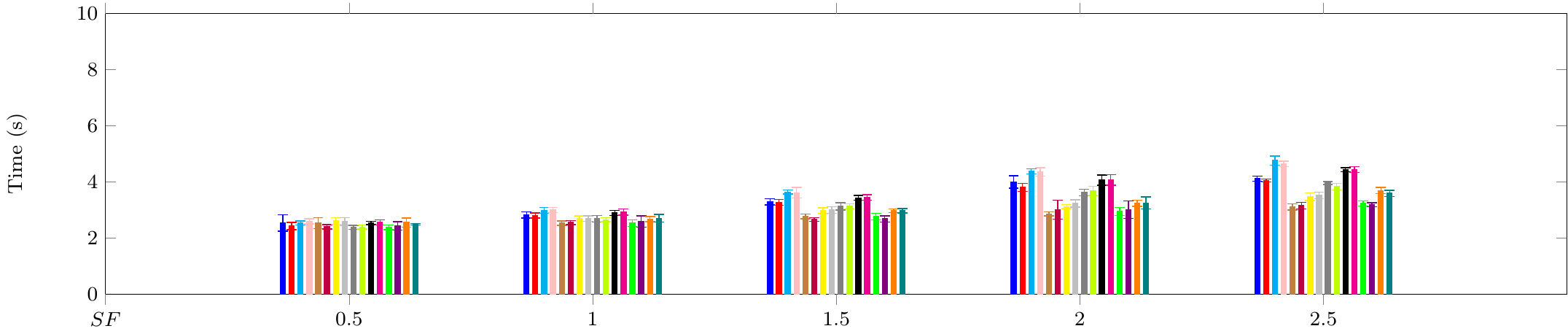}
		\caption{MongoDB MR top-$k$ documents: TF-IDF vs. Okapi BM25}
		\label{fig:mongo_tkd}
	\end{subfigure}
	\caption{Top-$k$ documents: TF-IDF vs. Okapi BM25 comparison}
	\label{fig:comp_tkd_tfidf_okapi}
\end{figure*}

Figure~\ref{fig:comp_tkk_tfidf_okapi} presents a performance comparison depending on the deployed distributed data management system and weighting scheme for retrieving top-$k$ keywords w.r.t. scale factor $SF$. Whereas Figure~\ref{fig:comp_tkd_tfidf_okapi} presents a similar performance comparison for top-$k$ documents.
These comparisons use only MongoDB's MR query implementation for both top-$k$ keywords and documents, as there is no AP implementation that uses Okapi BM25.

In the following paragraphs we analyze the results for each of the proposed distributed platforms.


\subsubsection{Hive} 

For Hive, computing the top-$k$ keywords with TF-IDF is faster than computing them with Okapi BM25 (Figure~\ref{fig:hive_tkk}) with a factor between $1.2$x and $1.4$x. This is applicable for all queries, regardless of $SF$. 
The difference in performance between the weighting schemes are almost the same for each query w.r.t. $SF$. The biggest difference in runtime between TF-IDF and Okapi BM25 is for query $Q_1$, while the smallest is obtained for query $Q_4$ because of its  higher selectivity. The difference in performance between the two weighting schemes is directly impacted by the query selectivity, complexity and the $SF$. Moreover, this difference is also impacted by the computational complexity of the weighting schemes. 

For calculating the top-$k$ documents, TF-IDF runtime is smaller than Okapi BM25 (Figure~\ref{fig:hive_tkd}). 
As in the case of top-$k$ keywords, the gap between the two weighting schemes increases with $SF$, which directly impacts each query's selectivity.
The biggest gap in runtime between TF-IDF and Okapi BM25 can be observed in the case of $Q'_1$ and $Q'_3$ queries, while the smallest is obtained for $Q'_2$ and $Q'_4$ queries.
In conclusion, the performance differences between the two weighting schemes are directly impacted by the weight computational complexity, the number of dimensions traversed by each query, and the $SF$.

\subsubsection{Spark}

For the Spark environment, computing top-$k$ keywords with TF-IDF in comparison with Okapi BM25 is faster (Figure~\ref{fig:spark_tkk}), although the differences in runtime performance are even lower than the Hive setup. 
The performance is directly influenced by the framework and by how fast the resources are allocated to the worker nodes by the YARN resource manager. Thus, resource allocation latency also increases the standard deviation measured for each experiment.

The same pattern is obtained when calculating the top-$k$ documents. It must be mentioned that in this case, using TF-IDF is faster than using Okapi BM25 with a factor of approximately $1.1$x (Figure~\ref{fig:spark_tkd}). Likewise, all queries have almost the same runtime differences, regardless of $SF$, complexity and selectivity. This is a direct impact of the application containers created by YARN, as they are created at the beginning of the application execution and the resources are fully allocated before running any job. Furthermore, Spark optimizes the computation through the Directed Acyclic Graph (DAG) execution engine that uses lazy evaluation for each tasks.

\subsubsection{MongoDB}

The last proposed scenario uses MongoDB and MapReduce as distributed platform. 

When computing the top-$k$ keywords, the runtime increases with a factor between $2$x and $3$x  for Okapi BM25 as opposed to TF-IDF. The largest performance gap is obtained for $Q_1$ and $Q_3$ queries, while the smallest is obtained for $Q_2$ and $Q_4$ queries. 
These performance outcomes are directly influenced by the intermediate "Sort and Shuffle" phase of the MapReduce algorithm. In this step all the results from all the Map functions are sorted and concatenated by key and sent to the Reducer functions to be aggregated. During this step, the shuffler component redistribute data based on the output keys which introduces additional computations, thus increasing the runtime.

When computing top-k documents with MongoDB, the runtime gap between the two schemes is greatly reduced. 
Although, the difference in execution times is small for the same $SF$ when computing top-$k$ documents with TF-IDF than Okapi BM25. 
These results are directly influenced by 
i) the queries' selectivity, i.e. the number of results returned by the to-$k$ documents queries is small compared to top-$k$ keywords,
and 
ii) MongoDB's schemaless and flexible data model, i.e., JOIN operations and labels with no information are eliminates using this model.

\subsection{Database Implementation Comparison}

The following set of experiments analysis the time performance of the different database implementations w.r.t. $SF$ and weighting schemes for the top-$k$ keywords and documents queries. 

\subsubsection{Top-$k$ keywords using TF-IDF}

\begin{figure}[!htbp]
	\centering
	\begin{subfigure}{0.9\columnwidth}
		\centering
		\includegraphics[width=\columnwidth]{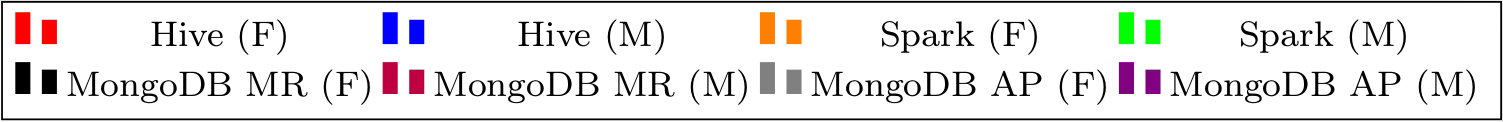}
	\end{subfigure}
	\begin{subfigure}{0.45\columnwidth}
		\centering
		\includegraphics[width=\columnwidth]{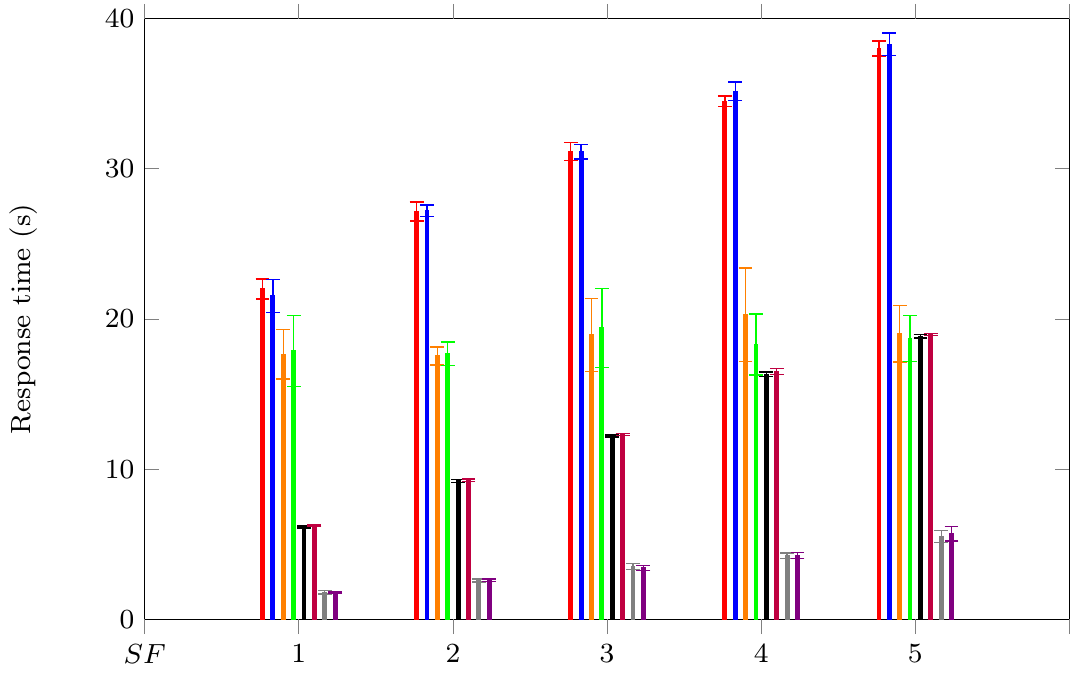}
		\caption{$Q_1$ TF-IDF}
		\label{fig:tkk_tfidf_q1}
	\end{subfigure}
	\begin{subfigure}{0.45\columnwidth}
		\centering
		\includegraphics[width=\columnwidth]{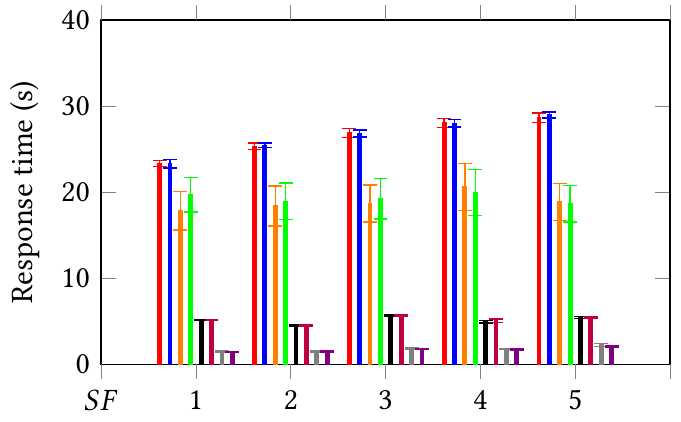}
		\caption{$Q_2$ TF-IDF}
		\label{fig:tkk_tfidf_q2}
	\end{subfigure}
	\begin{subfigure}{0.45\columnwidth}
		\centering
		\includegraphics[width=\columnwidth]{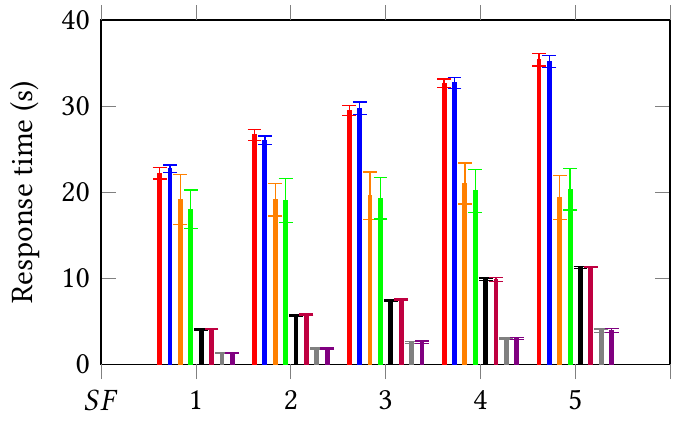}
		\caption{$Q_3$ TF-IDF}
		\label{fig:tkk_tfidf_q3}
	\end{subfigure}
	\begin{subfigure}{0.45\columnwidth}
		\centering
		\includegraphics[width=\columnwidth]{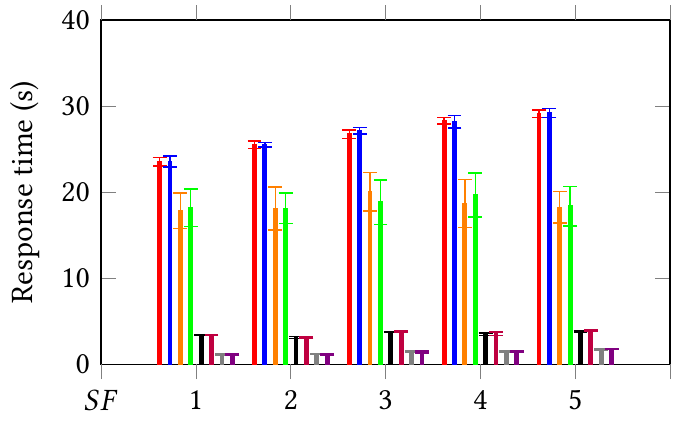}
		\caption{$Q_4$ TF-IDF}
		\label{fig:tkk_tfidf_q4}
	\end{subfigure}
	\caption{TF-IDF: Top-$k$ keywords query comparison}
	\label{fig:tkk_tfidf}
\end{figure}

Figure~\ref{fig:tkk_tfidf} presents the results' comparison for each query when computing the top-$k$ keywords with TF-IDF. 

Hive has the overall worst runtime for all queries. 
In comparison, Spark has a constant execution time regardless of $SF$. Whereas, with the increases of $SF$, the runtime gap decreases between Spark and Hive by a factor between $2$x and $1.8$x for all queries. This runtime gap can be explained by the fact that Hive uses Tez as query execution engine while Spark uses direct in-memory data processing. Tez is built on Hadoop MapReduce and relies extensively on HDFS, thus with the number of I/O operations the execution time increases. Spark's runtime performance is directly influenced by the chosen planning policy of the YARN resource manager, and implicitly by how the resources are allocated for each task.

The MongoDB distributed setup has the overall best runtime when using the TF-IDF weighting scheme for top-$k$ keywords.
For data aggregation, MongoDB provides a native aggregation framework, i.e, Aggregation Pipeline (AP), or the general framework MapReduce (MP). MP functionality offers more flexibility than the AP framework, but AP is optimized to work with data processing pipelines to increase query runtime performance. Thus, the best performance is obtained when using the MongoDB with AP. Whereas, when using MongoDB with MR, the time performance decreases by a factor of $2$ for $Q_2$ and $Q_4$ queries (Figures~\ref{fig:tkk_tfidf_q2} and~\ref{fig:tkk_tfidf_q4}) and by a factor of $3$ for  $Q_1$ and $Q_3$ queries (Figures~\ref{fig:tkk_tfidf_q1} and~\ref{fig:tkk_tfidf_q3}). 

Although MongoDB MR has a better execution time than Spark's, the gap disappears for query $Q_1$ and a larger $SF$ (Figure~\ref{fig:tkk_tfidf_q1}). 
This trend may be a consequence of the resource allocation policies used by the two systems.

\subsubsection{Top-$k$ keywords using Okapi BM25}
\label{sec:kkwBM25}

\begin{figure}[!htbp]
	\centering
	\begin{subfigure}{0.6\columnwidth}
		\centering
		\includegraphics[width=\columnwidth]{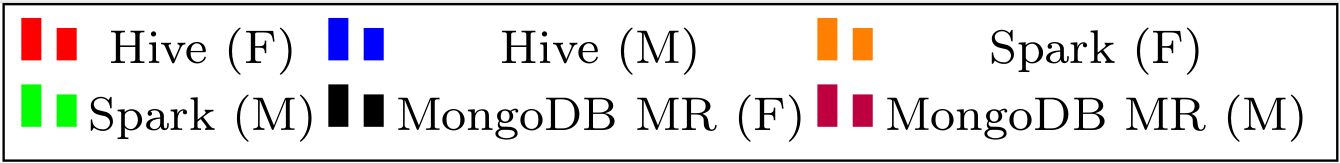}
	\end{subfigure}
	\begin{subfigure}{0.45\columnwidth}
		\centering
		\includegraphics[width=\columnwidth]{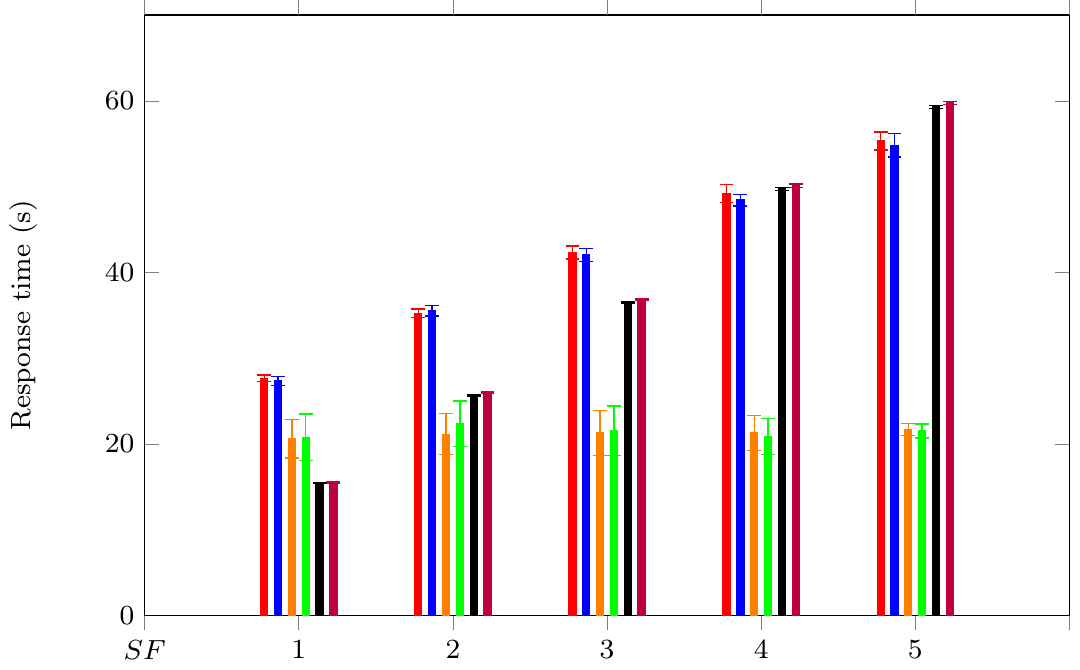}
		\caption{$Q_1$ Okapi BM25}
		\label{fig:tkk_okapi_q1}
	\end{subfigure}
	\begin{subfigure}{0.45\columnwidth}
		\centering
		\includegraphics[width=\columnwidth]{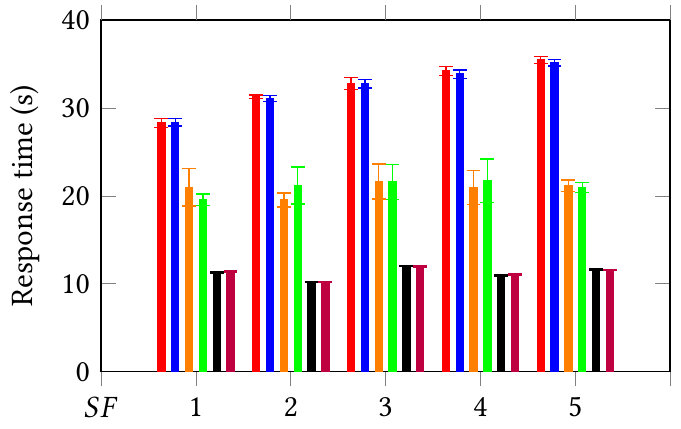}
		\caption{$Q_2$ Okapi BM25}
		\label{fig:tkk_okapi_q2}
	\end{subfigure}
	\begin{subfigure}{0.45\columnwidth}
		\centering
		\includegraphics[width=\columnwidth]{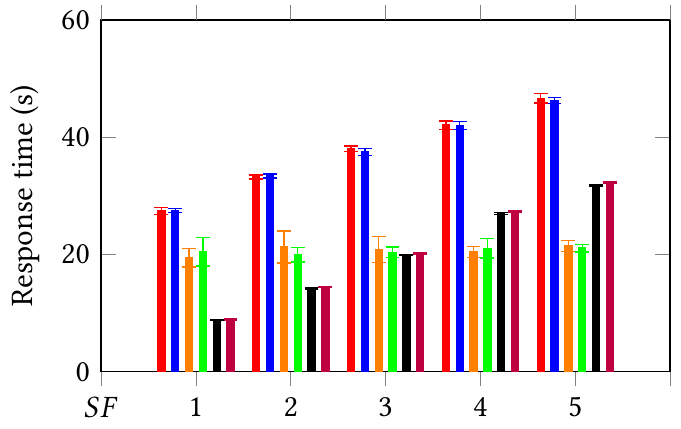}
		\caption{$Q_3$ Okapi BM25}
		\label{fig:tkk_okapi_q3}
	\end{subfigure}
	\begin{subfigure}{0.45\columnwidth}
		\centering
		\includegraphics[width=\columnwidth]{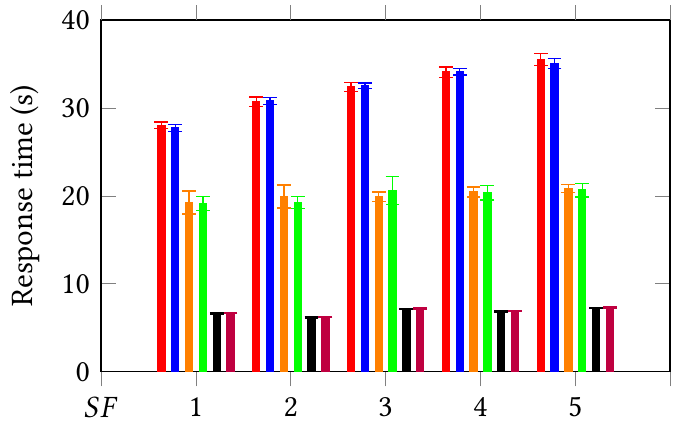}
		\caption{$Q_4$ Okapi BM25}
		\label{fig:tkk_okapi_q4}
	\end{subfigure}
	\caption{Okapi BM25: Top-$k$ keywords query comparison}
	\label{fig:tkk_okapi}
\end{figure}

Figure~\ref{fig:tkk_okapi} presents a comparison of the obtained results for each query when computing the top-$k$ keywords with Okapi BM25. 

For the MongoDB setup we use only the MR framework, as the Okapi BM25 implementation for the AP framework is not possible due to the complexity of implementing the nested queries used by the weighting function.

Hive has the worst performance for $Q_2$, $Q_3$ and $Q_4$ queries. Likewise, the performance slightly decreases for $Q_2$ (Figure~\ref{fig:tkk_okapi_q2}) and $Q_4$ (Figure~\ref{fig:tkk_okapi_q4}) queries, with the increase of $SF$.

For this set of experiments, Spark's runtime remains again constant. 

MongoDB obtains the overall best performance for $Q_2$ (Figure~\ref{fig:tkk_okapi_q2}) and $Q_4$ (Figure~\ref{fig:tkk_okapi_q4}) queries. For these two queries, the execution time is decreased by a factor of $2$x in comparison with Spark and by a factor of $4$x in comparison with Hive. Moreover, for these two queries the runtime is almost constant when changing the $SF$ factor.
As for $Q_1$ query, the runtime worsens with the increase of $SF$ (Figure~\ref{fig:tkk_okapi_q1}) to the point where it is lower than the performance of Hive. Ultimately, for $Q_3$ query with a small $SF$, MongoDB has the best execution time, but with increasing the $SF$ factor, the performance gets worse to the point that is outperformed by the Spark runtime. These results are a direct influence by the horizontal sharding policies used for distributing the data between the nodes.
We used as Sharding Key the unique record identifier, thus on some shards the distribution of data for the constrains is higher that on other shards. This distribution directly influences the workload on each node and ultimately the overall query performance. 

\subsubsection{Top-$k$ documents using TF-IDF}

\begin{figure}[!htbp]
	\centering
	\begin{subfigure}{0.6\columnwidth}
		\centering
		\includegraphics[width=\columnwidth]{Header_2.png}
	\end{subfigure}
	\begin{subfigure}{0.45\columnwidth}
		\centering
		\includegraphics[width=\columnwidth]{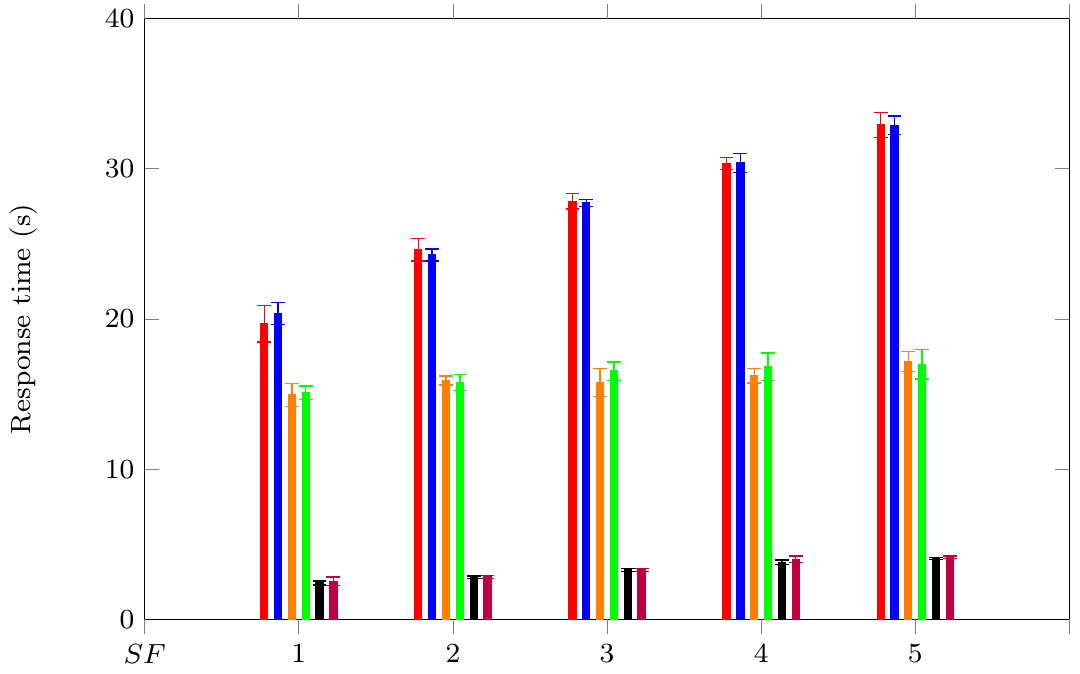}
		\caption{$Q'_1$ TF-IDF}
		\label{fig:tkd_tfidf_q1}
	\end{subfigure}
	\begin{subfigure}{0.45\columnwidth}
		\centering
		\includegraphics[width=\columnwidth]{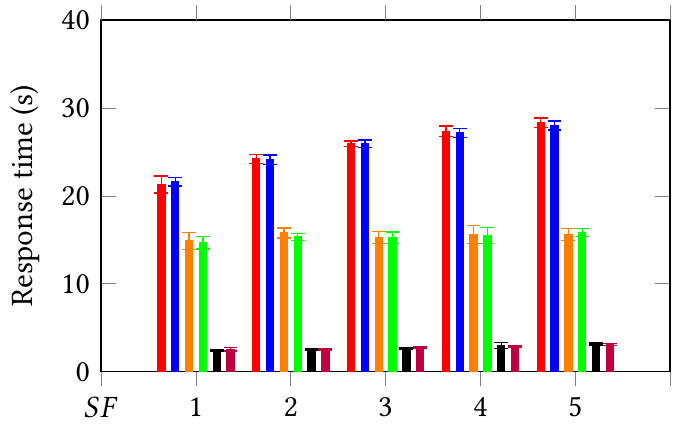}
		\caption{$Q'_2$ TF-IDF}
		\label{fig:tkd_tfidf_q2}
	\end{subfigure}
	\begin{subfigure}{0.45\columnwidth}
		\centering
		\includegraphics[width=\columnwidth]{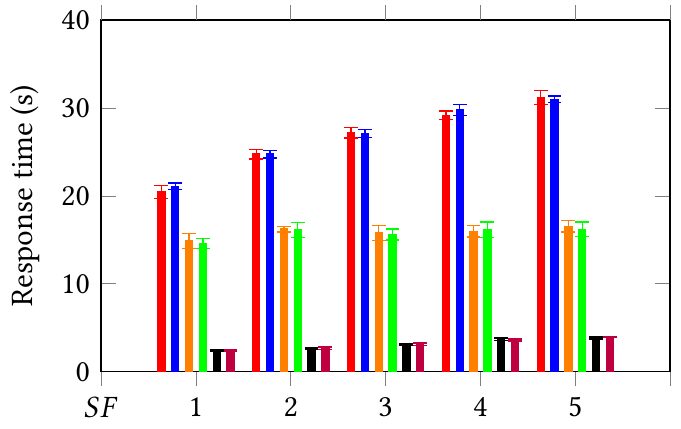}
		\caption{$Q'_3$ TF-IDF}
		\label{fig:tkd_tfidf_q3}
	\end{subfigure}
	\begin{subfigure}{0.45\columnwidth}
		\centering
		\includegraphics[width=\columnwidth]{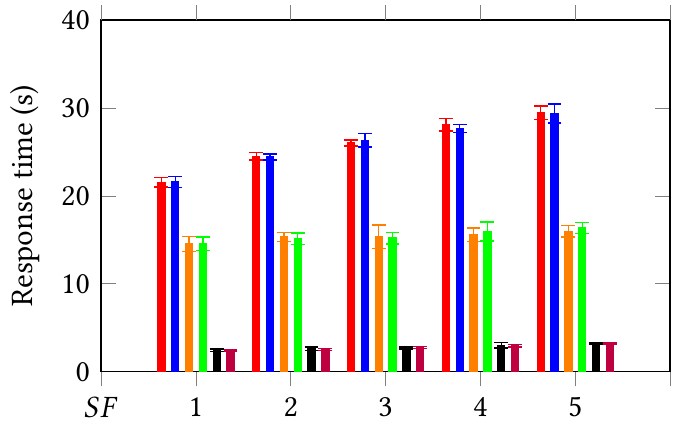}
		\caption{$Q'_4$ TF-IDF}
		\label{fig:tkd_tfidf_q4}
	\end{subfigure}
	\caption{TF-IDF: Top-$k$ documents query comparison}
	\label{fig:tkd_tfidf}
\end{figure}

The results obtained when computing the top-$k$ documents with TF-IDF are presented in Figure~\ref{fig:tkd_tfidf}.

Hive has the overall worst runtime between the tested systems. The decrease of performance while increasing the $SF$ factor follows the same pattern for all queries.  Likewise, by increasing $SF$, the execution time decreases by a factor of $2$x in comparison with Spark.

Spark's runtime is again almost constant for all queries. 

In comparison with the other systems, MongoDB has the best overall runtime, as well as maintaining a constant execution time. Moreover, with the increase of $SF$, the execution time decreases by a factor of $4$x in comparison with Spark and by a factor of $8$x in comparison with Hive. As in the previous set of tests (Subsection~\ref{sec:kkwBM25}), this is a direct impact of the sharding policies.

\subsubsection{Top-$k$ documents using Okapi BM25}

\begin{figure}[!htbp]
	\centering
	\begin{subfigure}{0.6\columnwidth}
		\centering
		\includegraphics[width=\columnwidth]{Header_2.png}
	\end{subfigure}
	\begin{subfigure}{0.45\columnwidth}
		\centering
		\includegraphics[width=\columnwidth]{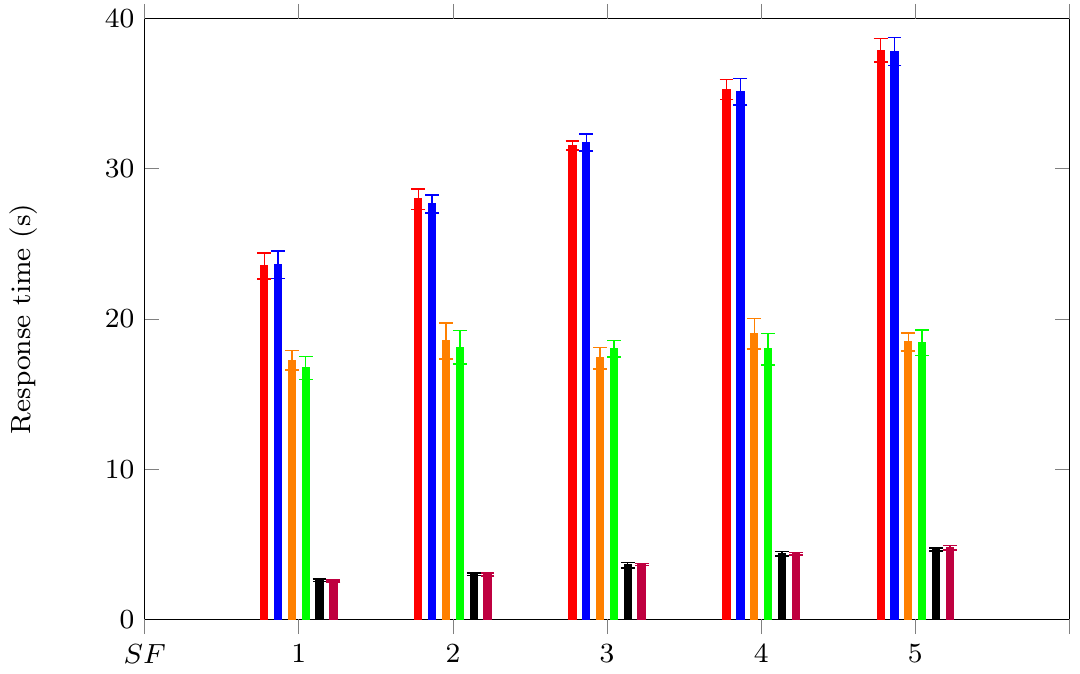}
		\caption{$Q'_1$ Okapi BM25}
		\label{fig:tkd_okapi_q1}
	\end{subfigure}
	\begin{subfigure}{0.45\columnwidth}
		\centering
		\includegraphics[width=\columnwidth]{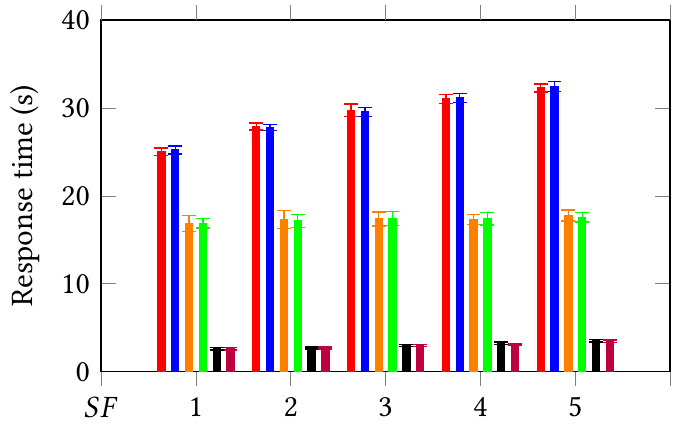}
		\caption{$Q'_2$ Okapi BM25}
		\label{fig:tkd_okapi_q2}
	\end{subfigure}
	\begin{subfigure}{0.45\columnwidth}
		\centering
		\includegraphics[width=\columnwidth]{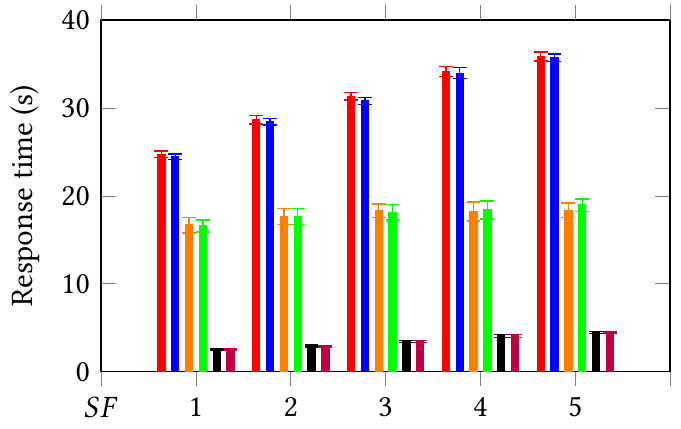}
		\caption{$Q'_3$ Okapi BM25}
		\label{fig:tkd_okapi_q3}
	\end{subfigure}
	\begin{subfigure}{0.45\columnwidth}
		\centering
		\includegraphics[width=\columnwidth]{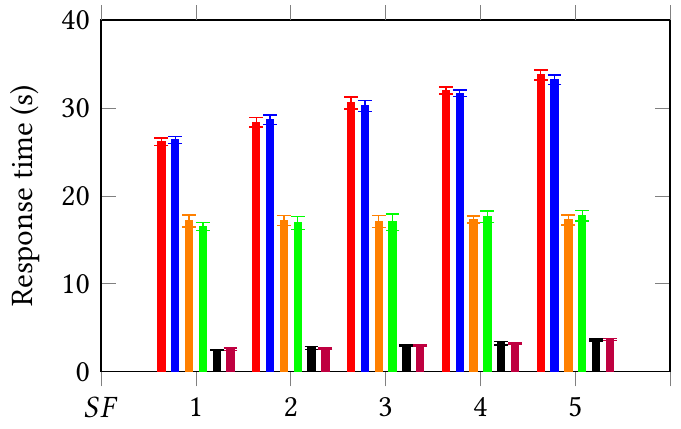}
		\caption{$Q'_4$ Okapi BM25}
		\label{fig:tkd_okapi_q4}
	\end{subfigure}
	\caption{Okapi BM25: Top-$k$ documents query comparison}
	\label{fig:tkd_okapi}
\end{figure}

The same runtime patterns emerge when computing the top-$k$ documents with Okapi BM25 as in the case of using the TF-IDF weighting scheme. The results are presented in Figure~\ref{fig:tkd_okapi}.


Hive has again the overall worst performance. The execution time decreases by a factor of $2$x in comparison with Spark while increasing $SF$ for all queries.

Likewise, we obtain the same pattern for Spark's runtime. The overall execution time is almost constant regardless of $SF$. This constant performance is due to the initialization step of the Spark Context. Moreover, the resource allocation done by YARN also influence the runtime performance.

The best overall performance is again achieved by MongoDB. Moreover, we observe that with increasing $SF$, the execution time is almost constant for all queries. This constant performance is influenced by the distribution of the data on shards and how the MapReduce tasks handles them. 

\section{Conclusion}\label{sec:conclusions}

In this paper, we present our distributed benchmark solution for any type of textual data -- \textsc{TextBenDS}.
Our benchmark tests the computation efficiency of term weighting schemes. These weights are computed using data sampling methods and aggregation queries. Data sampling enables analysis based on gender, location, and time to extract general linguistic and social context features, while aggregation queries compute dynamically the weighting schemes and ranking functions. The experiments done on different distributed systems prove that \textsc{TextBenDS} is adequate for structured, semi-structured, and unstructured textual data. Furthermore, our solution proves its portability, scalability, and relevance by its design.

We proved that our solution is portable, as it works on multiple distributed systems. For this purpose, we compare the performance of several such systems, e.g. Hive - a distributed DBMS, Spark - a distributed framework using a relational approach, and MongoDB - a document-oriented DBMS. \textsc{TextBenDS} tests employ two different weighting schemes (TF-IDF and Okapi BM25) for processing top-$k$ keywords and documents. 

To demonstrate the scalability of our solution, we introduced $SF$, the scaling factor that introduces an incremental growth in the data volume for our experiments. 
The queries complexity together with $SF$ produces a linear increase in runtime for Hive and MongoDB, while Sparks displays a constant execution time. 

Another important property of \textsc{TextBenDS} is its relevance in performance analysis and text mining. This is proven by the fact that our solution analysis the runtime performance for ranking keywords and documents, techniques widely used in various text analystics and retrieval tasks.


As expected, computing the top-$k$ keywords with TF-IDF is generally faster than with Okapi BM25, regardless of the deployed distributed system.
However, the difference in performance between TF-IDF and Okapi BM25 diverge from one system to another. When using Hive, this performance gap between the two weighting schemes is between $1.2$x and $1.4$x. Instead, for Spark the gap is diminishing up to a factor of $1.1$x or less. This is largely due to how fast the resources are allocated to the worker nodes by the Spark's resource manager.
However, the largest performance gap for computing the top-$k$ keywords is obtained with MongoDB and MapReduce framework. In this case the runtime increases with a factor between $2$x and $3$x. We deduced that these performance outcomes are directly influenced by the intermediate "Sort and Shuffle" phase of the MapReduce algorithm.

For the next set of experiments, focused on computing the top-$k$ documents, the performance gap between TF-IDF and Okapi BM25 is decreasing for all of the deployed distributed systems. The smallest difference is found in running with Spark and MongoDB, where the runtime gap reaches a factor of $1.1$x, and even $1$x. 

As an analysis of the overall performance of the three distributed systems, it resulted that the best outcomes are obtained in some cases by MongoDB with MapReduce, and in other cases by Spark. 

Although we evaluate \textsc{TextBenDS} on social media data, our benchmark is generic and can be used on any textual data enhanced with metadata extracted through preprocessing and Natural Language Processing techniques, e.g., part-of speech tagging, lemmatization, hashtag extraction, etc.



As future work, we plan to improve the scalability of our solution by designing sampling strategies and aggregation queries. 
The sampling methods will include constraints on tags and named entities while boosting the performance by lowering the query selectivity and complexity. Furthermore, we plan to expand \textsc{TextBenDS}'s dataset significantly to achieve a big data-scale volume.
Moreover, we considered in this paper TF-IDF and Okapi BM25 for machine learning tasks. The next version of our benchmark should include other weighting schemes, such as KL-divergence~\cite{Raiber2017}, to further improve the relevance of our solution. 



\bibliographystyle{spmpsci}
\bibliography{textbends} 

\begin{thebibliography}{10}
\providecommand{\url}[1]{{#1}}
\providecommand{\urlprefix}{URL }
\expandafter\ifx\csname urlstyle\endcsname\relax
  \providecommand{\doi}[1]{DOI~\discretionary{}{}{}#1}\else
  \providecommand{\doi}{DOI~\discretionary{}{}{}\begingroup
  \urlstyle{rm}\Url}\fi

\bibitem{Agrawal2016}
Agrawal, D., Butt, A., Doshi, K., Larriba-Pey, J.L., Li, M., Reiss, F.R., Raab,
  F., Schiefer, B., Suzumura, T., Xia, Y.: Sparkbench -- a spark performance
  testing suite.
\newblock In: Performance Evaluation and Benchmarking: Traditional to Big Data
  to Internet of Things, pp. 26--44. Springer International Publishing (2016).
\newblock \doi{10.1007/978-3-319-31409-9_3}

\bibitem{Armbrust2015}
Armbrust, M., Xin, R.S., Lian, C., Huai, Y., Liu, D., Bradley, J.K., Meng, X.,
  Kaftan, T., Franklin, M.J., Ghodsi, A., Zaharia, M.: Spark sql: Relational
  data processing in spark.
\newblock In: ACM SIGMOD International Conference on Management of Data, pp.
  1383--1394. {ACM} Press (2015).
\newblock \doi{10.1145/2723372.2742797}

\bibitem{Armstrong2013}
Armstrong, T.G., Ponnekanti, V., Borthakur, D., Callaghan, M.: Linkbench: A
  database benchmark based on the facebook social graph.
\newblock In: Proceedings of the 2013 ACM SIGMOD International Conference on
  Management of Data, SIGMOD '13, pp. 1185--1196. ACM (2013).
\newblock \doi{10.1145/2463676.2465296}

\bibitem{Bellot2013}
Bellot, P., Doucet, A., Geva, S., Gurajada, S., Kamps, J., Kazai, G., Koolen,
  M., Mishra, A., Moriceau, V., Mothe, J., Preminger, M., SanJuan, E.,
  Schenkel, R., Tannier, X., Theobald, M., Trappett, M., Trotman, A.,
  Sanderson, M., Scholer, F., Wang, Q.: Report on inex 2013.
\newblock SIGIR Forum \textbf{47}(2), 21--32 (2013).
\newblock \doi{10.1145/2568388.2568393}

\bibitem{Bifet2010}
Bifet, A., Frank, E.: Sentiment knowledge discovery in twitter streaming data.
\newblock In: Discovery Science, pp. 1--15. Springer Berlin Heidelberg (2010).
\newblock \doi{10.1007/978-3-642-16184-1_1}

\bibitem{Bouakkaz2016}
Bouakkaz, M., Loudcher, S., Ouinten, Y.: {OLAP} textual aggregation approach
  using the google similarity distance.
\newblock International Journal of Business Intelligence and Data Mining
  \textbf{11}(1), 31 (2016).
\newblock \doi{10.1504/ijbidm.2016.076425}

\bibitem{Bringay2011}
Bringay, S., B{\'e}chet, N., Bouillot, F., Poncelet, P., Roche, M., Teisseire,
  M.: Towards an on-line analysis of tweets processing.
\newblock In: International Conference on Database and Expert Systems
  Applications (DEXA), pp. 154--161 (2011).
\newblock \doi{10.1007/978-3-642-23091-2_15}

\bibitem{Chowdhury2014}
Chowdhury, B., Rabl, T., Saadatpanah, P., Du, J., Jacobsen, H.A.: A bigbench
  implementation in the hadoop ecosystem.
\newblock In: Advancing Big Data Benchmarks, pp. 3--18. Springer International
  Publishing (2014).
\newblock \doi{10.1007/978-3-319-10596-3_1}

\bibitem{Crane2017}
Crane, M., Culpepper, J.S., Lin, J., Mackenzie, J., Trotman, A.: A comparison
  of document-at-a-time and score-at-a-time query evaluation.
\newblock In: Proceedings of the Tenth ACM International Conference on Web
  Search and Data Mining, pp. 201--210. ACM (2017).
\newblock \doi{10.1145/3018661.3018726}

\bibitem{Dean2008}
Dean, J., Ghemawat, S.: Mapreduce: Simplified data processing on large
  clusters.
\newblock Communications of the ACM \textbf{51}(1), 107--113 (2008).
\newblock \doi{10.1145/1327452.1327492}

\bibitem{Deerwester1990}
Deerwester, S., Dumais, S.T., Furnas, G.W., Landauer, T.K., Harshman, R.:
  Indexing by latent semantic analysis.
\newblock Journal of the American Society for Information Science
  \textbf{41}(6), 391--407 (1990).
\newblock \doi{10.1002/(SICI)1097-4571(199009)41:6<391::AID-ASI1>3.0.CO;2-9}

\bibitem{Ferrarons2014}
Ferrarons, J., Adhana, M., Colmenares, C., Pietrowska, S., Bentayeb, F.,
  Darmont, J.: Primeball: a parallel processing framework benchmark for big
  data applications in the cloud.
\newblock In: 5th TPC Technology Conference on Performance Evaluation and
  Benchmarking (TPCTC 2013), \emph{LNCS1}, vol. 839, pp. 109--124 (2014).
\newblock \doi{10.1007/978-3-319-04936-6_8}

\bibitem{Gattiker2013}
Gattiker, A.E., Gebara, F.H., Hofstee, H.P., Hayes, J.D., Hylick, A.: Big data
  text-oriented benchmark creation for {Hadoop}.
\newblock {IBM} Journal of Research and Development \textbf{57}(3/4),
  10:1--10:6 (2013).
\newblock \doi{10.1147/JRD.2013.2240732}

\bibitem{Ghazal2017}
{Ghazal}, A., {Ivanov}, T., {Kostamaa}, P., {Crolotte}, A., {Voong}, R.,
  {Al-Kateb}, M., {Ghazal}, W., {Zicari}, R.V.: Bigbench v2: The new and
  improved bigbench.
\newblock In: 2017 IEEE 33rd International Conference on Data Engineering
  (ICDE), pp. 1225--1236 (2017).
\newblock \doi{10.1109/ICDE.2017.167}

\bibitem{Ghazal2013}
Ghazal, A., Rabl, T., Hu, M., Raab, F., Poess, M., Crolotte, A., Jacobsen,
  H.A.: Bigbench: Towards an industry standard benchmark for big data
  analytics.
\newblock In: Proceedings of the 2013 ACM SIGMOD International Conference on
  Management of Data, SIGMOD '13, pp. 1197--1208 (2013).
\newblock \doi{10.1145/2463676.2463712}

\bibitem{Gray1993}
Gray, J.: The Benchmark Handbook for Database and Transaction Systems (2nd
  Edition).
\newblock Morgan Kaufmann (1993)

\bibitem{Guille2015}
Guille, A., Favre, C.: Event detection, tracking, and visualization in twitter:
  a mention-anomaly-based approach.
\newblock Social Network Analysis and Mining \textbf{5}(1), 18 (2015).
\newblock \doi{10.1007/s13278-015-0258-0}

\bibitem{Hofmann2017}
Hofmann, T.: Probabilistic latent semantic indexing.
\newblock SIGIR Forum \textbf{51}(2), 211--218 (2017).
\newblock \doi{10.1145/3130348.3130370}

\bibitem{Huang2010}
Huang, S., Huang, J., Dai, J., Xie, T., Huang, B.: {The HiBench benchmark
  suite: Characterization of the MapReduce-based data analysis}.
\newblock In: Workshops Proceedings of the 26th International Conference on
  Data Engineering ({ICDE} 2010), pp. 41--51 (2010).
\newblock \doi{10.1109/ICDEW.2010.5452747}

\bibitem{Jia2014}
{Jia}, Z., {Zhan}, J., {Wang}, L., {Han}, R., {McKee}, S.A., {Yang}, Q., {Luo},
  C., {Li}, J.: Characterizing and subsetting big data workloads.
\newblock In: 2014 IEEE International Symposium on Workload Characterization
  (IISWC), pp. 191--201 (2014).
\newblock \doi{10.1109/IISWC.2014.6983058}

\bibitem{Krasnashchok2018}
Krasnashchok, K., Jouili, S.: Improving topic quality by promoting named
  entities in topic modeling.
\newblock In: Annual Meeting of the Association for Computational Linguistics,
  pp. 247--253 (2018)

\bibitem{Kilinc2017}
Kılın{\c{c}}, D., {\"{O}}z{\c{c}}ift, A., Bozyigit, F., Yildirim, P.,
  Y{\"{u}}calar, F., Borandag, E.: Ttc-3600: A new benchmark dataset for
  turkish text categorization.
\newblock Journal of Information Science \textbf{43}(2), 174--185 (2017).
\newblock \doi{10.1177/0165551515620551}

\bibitem{Lavrenko2017}
Lavrenko, V., Croft, W.B.: Relevance-based language models.
\newblock SIGIR Forum \textbf{51}(2), 260--267 (2017).
\newblock \doi{10.1145/3130348.3130376}

\bibitem{Lewis2004}
Lewis, D.D., Yang, Y., Rose, T.G., Li, F.: Rcv1: A new benchmark collection for
  text categorization research.
\newblock Journal of Machine Learning Research \textbf{5}, 361--397 (2004).
\newblock \urlprefix\url{http://www.jmlr.org/papers/v5/lewis04a.html}

\bibitem{Li2015}
Li, M., Tan, J., Wang, Y., Zhang, L., Salapura, V.: Sparkbench: A comprehensive
  benchmarking suite for in memory data analytic platform spark.
\newblock In: Proceedings of the 12th ACM International Conference on Computing
  Frontiers, CF '15, pp. 53:1--53:8. ACM (2015).
\newblock \doi{10.1145/2742854.2747283}

\bibitem{Lin2016}
Lin, J., Crane, M., Trotman, A., Callan, J., Chattopadhyaya, I., Foley, J.,
  Ingersoll, G., Macdonald, C., Vigna, S.: Toward reproducible baselines: The
  open-source ir reproducibility challenge.
\newblock In: Advances in Information Retrieval, pp. 408--420. Springer
  International Publishing (2016).
\newblock \doi{10.1007/978-3-319-30671-1_30}

\bibitem{Manning2008}
Manning, C.D., Raghavan, P., Sch{\"u}tze, H.: Introduction to information
  retrieval.
\newblock Cambridge University Press (2008)

\bibitem{Ming2014}
Ming, Z., Luo, C., Gao, W., Han, R., Yang, Q., Wang, L., Zhan, J.: Bdgs: A
  scalable big data generator suite in big data benchmarking.
\newblock In: Advancing Big Data Benchmarks, pp. 138--154. Springer
  International Publishing (2014).
\newblock \doi{10.1007/978-3-319-10596-3_11}

\bibitem{OShea2010}
O'Shea, J., Bandar, Z., Crockett, K.A., McLean, D.: Benchmarking short text
  semantic similarity.
\newblock International Journal of Intelligent Information and Database Systems
  \textbf{4}(2), 103--120 (2010).
\newblock \doi{10.1504/IJIIDS.2010.032437}

\bibitem{Paltoglou2010}
Paltoglou, G., Thelwall, M.: A study of information retrieval weighting schemes
  for sentiment analysis.
\newblock In: 48th Annual Meeting of the Association for Computational
  Linguistics, pp. 1386--1395 (2010).
\newblock \urlprefix\url{http://dl.acm.org/citation.cfm?id=1858681.1858822}

\bibitem{Partalas2015}
Partalas, I., Kosmopoulos, A., Baskiotis, N., Arti{\`{e}}res, T., Paliouras,
  G., Gaussier, {\'{E}}., Androutsopoulos, I., Amini, M.R., Gallinari, P.:
  Lshtc: A benchmark for large-scale text classification.
\newblock CoRR  (2015).
\newblock \urlprefix\url{http://arxiv.org/abs/1503.08581}

\bibitem{Pirzadeh2015}
{Pirzadeh}, P., {Carey}, M.J., {Westmann}, T.: Bigfun: A performance study of
  big data management system functionality.
\newblock In: 2015 IEEE International Conference on Big Data (Big Data), pp.
  507--514 (2015).
\newblock \doi{10.1109/BigData.2015.7363793}

\bibitem{Raiber2017}
Raiber, F., Kurland, O.: Kullback-leibler divergence revisited.
\newblock In: Proceedings of the ACM SIGIR International Conference on Theory
  of Information Retrieval, ICTIR '17, pp. 117--124. ACM (2017).
\newblock \doi{10.1145/3121050.3121062}

\bibitem{Ravat2008}
Ravat, F., Teste, O., Tournier, R., Zurfluh, G.: Top\_keyword: an aggregation
  function for textual document olap.
\newblock In: 10th International Conference on Data Warehousing and Knowledge
  Discovery (DaWaK), pp. 55--64 (2008).
\newblock \doi{10.1007/978-3-540-85836-2_6}

\bibitem{Saha2015}
Saha, B., Shah, H., Seth, S., Vijayaraghavan, G., Murthy, A., Curino, C.:
  Apache tez: A unifying framework for modeling and building data processing
  applications.
\newblock In: ACM SIGMOD International Conference on Management of Data, pp.
  1357--1369. ACM, New York, NY, USA (2015).
\newblock \doi{10.1145/2723372.2742790}

\bibitem{Sangroya2013}
Sangroya, A., Serrano, D., Bouchenak, S.: Mrbs: Towards dependability
  benchmarking for hadoop mapreduce.
\newblock In: Euro-Par 2012: Parallel Processing Workshops, pp. 3--12. Springer
  Berlin Heidelberg (2013).
\newblock \doi{10.1007/978-3-642-36949-0_2}

\bibitem{Shu2018}
Shu, K., Mahudeswaran, D., Wang, S., Lee, D., Liu, H.: Fakenewsnet: A data
  repository with news content, social context and dynamic information for
  studying fake news on social media.
\newblock arXiv preprint arXiv:1809.01286  (2018)

\bibitem{Shu2017}
Shu, K., Sliva, A., Wang, S., Tang, J., Liu, H.: Fake news detection on social
  media: A data mining perspective.
\newblock ACM SIGKDD Explorations Newsletter \textbf{19}(1), 22--36 (2017).
\newblock \doi{10.1145/3137597.3137600}

\bibitem{Shvachko2010}
Shvachko, K., Kuang, H., Radia, S., Chansler, R.: The hadoop distributed file
  system.
\newblock In: Symposium on Mass Storage Systems and Technologies, pp. 1--10
  (2010).
\newblock \doi{10.1109/MSST.2010.5496972}

\bibitem{SparckJones2000a}
{Spärck Jones}, K., Walker, S., Robertson, S.E.: A probabilistic model of
  information retrieval: development and comparative experiments: Part 1.
\newblock Information Processing \& Management \textbf{36}(6), 779 -- 808
  (2000).
\newblock \doi{10.1016/S0306-4573(00)00015-7}

\bibitem{SparckJones2000b}
{Spärck Jones}, K., Walker, S., Robertson, S.E.: A probabilistic model of
  information retrieval: development and comparative experiments: Part 2.
\newblock Information Processing \& Management \textbf{36}(6), 809 -- 840
  (2000).
\newblock \doi{10.1016/S0306-4573(00)00016-9}

\bibitem{Thusoo2009}
Thusoo, A., Sarma, J.S., Jain, N., Shao, Z., Chakka, P., Anthony, S., Liu, H.,
  Wyckoff, P., Murthy, R.: Hive: A warehousing solution over a map-reduce
  framework.
\newblock VLDB Endowment \textbf{2}(2), 1626--1629 (2009).
\newblock \doi{10.14778/1687553.1687609}

\bibitem{tpcxhs15}
{Transaction Processing Performance Council (TPC)}: {TPC Express Benchmark HS
  Standard Specification Version 1.4.2} (2016).
\newblock \urlprefix\url{http://www.tpc.org}

\bibitem{tpcds}
{Transaction Processing Performance Council (TPC)}: {TPC-DS Decision Support
  Benchmark 2.10.1} (2019).
\newblock \urlprefix\url{http://www.tpc.org}

\bibitem{Truica2016}
Truica, C.O., Radulescu, F., Boicea, A.: Comparing different term weighting
  schemas for topic modeling.
\newblock In: 2016 18th International Symposium on Symbolic and Numeric
  Algorithms for Scientific Computing ({SYNASC}). {IEEE} (2016).
\newblock \doi{10.1109/synasc.2016.055}

\bibitem{Truica2017}
Truică, C.O., Darmont, J.: {T$^2$K$^2$}: The twitter top-k keywords benchmark.
\newblock In: Communications in Computer and Information Science, pp. 21--28.
  Springer International Publishing (2017).
\newblock \doi{10.1007/978-3-319-67162-8_3}

\bibitem{Truica2018}
Truică, C.O., Darmont, J., Boicea, A., Rădulescu, F.: Benchmarking top-k
  keyword and top-k document processing with {T$^2$K$^2$} and
  {T$^2$K$^2$D$^2$}.
\newblock Future Generation Computer Systems \textbf{85}, 60--75 (2018).
\newblock \doi{10.1016/j.future.2018.02.037}

\bibitem{Vavilapalli2013}
Vavilapalli, V.K., Murthy, A.C., Douglas, C., Agarwal, S., Konar, M., Evans,
  R., Graves, T., Lowe, J., Shah, H., Seth, S., Saha, B., Curino, C., O'Malley,
  O., Radia, S., Reed, B., Baldeschwieler, E.: Apache hadoop yarn: Yet another
  resource negotiator.
\newblock In: Annual Symposium on Cloud Computing, pp. 5:1--5:16 (2013).
\newblock \doi{10.1145/2523616.2523633}

\bibitem{Wang2016}
Wang, L., Dong, X., Zhang, X., Wang, Y., Ju, T., Feng, G.: Textgen: a realistic
  text data content generation method for modern storage system benchmarks.
\newblock Frontiers of Information Technology \& Electronic Engineering
  \textbf{17}(10), 982--993 (2016).
\newblock \doi{10.1631/FITEE.1500332}

\bibitem{Wang2014}
Wang, L., Zhan, J., Luo, C., Zhu, Y., Yang, Q., He, Y., Gao, W., Jia, Z., Shi,
  Y., Zhang, S., Zheng, C., Lu, G., Zhan, K., Li, X., Qiu, B.: {BigDataBench:
  {A} big data benchmark suite from internet services}.
\newblock In: 20th {IEEE} International Symposium on High Performance Computer
  Architecture ({HPCA} 2014), pp. 488--499 (2014).
\newblock \doi{10.1109/HPCA.2014.6835958}

\bibitem{Wang2017}
{Wang}, X., {Ah-Pine}, J., {Darmont}, J.: Shcoclust, a scalable
  similarity-based hierarchical co-clustering method and its application to
  textual collections.
\newblock In: 2017 IEEE International Conference on Fuzzy Systems (FUZZ-IEEE),
  pp. 1--6 (2017).
\newblock \doi{10.1109/FUZZ-IEEE.2017.8015720}

\bibitem{Yin2018}
Yin, J., Chao, D., Liu, Z., Zhang, W., Yu, X., Wang, J.: Model-based clustering
  of short text streams.
\newblock In: ACM SIGKDD International Conference on Knowledge Discovery {\&}
  Data Mining, pp. 2634--2642. {ACM} Press (2018).
\newblock \doi{10.1145/3219819.3220094}

\bibitem{Zaharia2016}
Zaharia, M., Xin, R.S., Wendell, P., Das, T., Armbrust, M., Dave, A., Meng, X.,
  Rosen, J., Venkataraman, S., Franklin, M.J., Ghodsi, A., Gonzalez, J.,
  Shenker, S., Stoica, I.: Apache spark: A unified engine for big data
  processing.
\newblock Communications of the ACM \textbf{59}(11), 56--65 (2016).
\newblock \doi{10.1145/2934664}

\bibitem{Zhang2009}
Zhang, D., Zhai, C., Han, J.: Topic cube: Topic modeling for {OLAP} on
  multidimensional text databases.
\newblock In: Proceedings of the 2009 {SIAM} International Conference on Data
  Mining, pp. 1124--1135. Society for Industrial and Applied Mathematics
  (2009).
\newblock \doi{10.1137/1.9781611972795.96}

\bibitem{Zhang2012}
Zhang, D., Zhai, C., Han, J.: {MiTexCube}: {MicroTextCluster} cube for online
  analysis of text cells and its applications.
\newblock Statistical Analysis and Data Mining \textbf{6}(3), 243--259 (2012).
\newblock \doi{10.1002/sam.11159}

\end{thebibliography}

\end{document}